\documentclass[a4paper,UKenglish,cleveref, cleveref, thm-restate]{lipics-v2021}
\nolinenumbers
\hideLIPIcs

\title{Blockchain Nodes are Heterogeneous and Your P2P Overlay Should be Too: PODS } 

\author{Naqib Zarin}{Anoma \& Delft University of Technology, Amsterdam, The Netherlands}{naqib@heliax.dev}{}{}
\author{Isaac Sheff}{Anoma, Buffalo, USA}{isaac@heliax.dev}{}{}
\author{Stefanie Roos}{Delft University of Technology, Delft, The Netherlands}{s.roos@tudelft.nl}{}{}

\authorrunning{Naqib Zarin, Isaac Sheff, and Stefanie Roos} 



\usepackage[ruled]{algorithm2e}
\begin{document}

\maketitle

\begin{abstract}
At the core of each blockchain system, parties communicate through a peer-to-peer (P2P) overlay. Unfortunately, recent evidence suggests these P2P overlays represent a significant bottleneck for transaction throughput and scalability. Furthermore, they enable a number of attacks.  We argue that these performance and security problems arise because current P2P overlays cannot fully capture the complexity of a blockchain system as they do not offer flexibility to accommodate node heterogeneity. 

We propose a novel approach to address these issues: P2P Overlay Domains with Sovereignty (PODS), which allows nodes in a single overlay to belong to multiple heterogeneous groups, called domains. Each domain features its own set of protocols, tailored to the characteristics and needs of its nodes. To demonstrate the effectiveness of PODS, we design and implement two novel node discovery protocols: FedKad and SovKad. Using a custom simulator, we show that node discovery using PODS (SovKad) architecture outperforms both single overlay (Kademlia) and multi-overlay (FedKad) architectures in terms of hop count and success rate, though FedKad requires slightly less bandwidth.  
\end{abstract}

\section{Introduction}
\label{sec:introduction}

At the core of each blockchain system is the \textit{peer-to-peer P2P network}, which facilitates communication between nodes. Blockchain systems use a P2P architecture as a base due to their decentralized and self-stabilizing nature~\cite{zheng2017overview, surveynetworking}. Because nodes in the network communicate with each other directly, there is no need for a central server or entity to mediate between nodes~\cite{shen2010handbook}. 

The P2P network in a blockchain system is commonly a \textit{P2P overlay}: a virtual network built on top of the underlying physical infrastructure (i.e., the Internet) and responsible for managing virtual connections between nodes and ensuring that information is propagated efficiently throughout the network~\cite{shen2010handbook}.

Recent evidence suggests that the P2P overlay is the source of critical challenges such as low transaction throughput and poor security~\cite{9524210, dotan2021survey, narwhal}, which are currently major factors holding back adoption of blockchain technology~\cite{dotan2021survey}. 


Existing solutions primarily focus on temporary fixes within current overlays \cite{kadcast, bi2018accelerated, liu2022asymptotically}, but the limitations of the underlying overlays make these solutions ill-suited for sustainable long-term deployment. These overlays (e.g., \cite{kademlia, morris2003chord, srivatsa2006large}) were originally designed for applications that did not consider the inherent differences between nodes \cite{tarkoma2010overlay}, resulting in performance deficiencies in heterogeneous environments \cite{lopez2019please}. Blockchains, by their very nature, encompass heterogeneity: subsets of nodes possess different capabilities and preferences for latency, bandwidth, and failure tolerance trade-offs. For instance, light clients demand less bandwidth compared to validators as they frequently engage in one-to-one communication, while validator nodes participate in one-to-many information gossiping throughout the overlay network, and may be willing to use a lot of bandwidth if they can achieve low latency. An overlay architecture capable of accommodating node heterogeneity is indispensable to cater to these diverse requirements.

New P2P overlay architectures have emerged to accommodate node heterogeneity. Typically, they group nodes  into separate homogeneous sub-networks that require less communication with other sub-networks, and each sub-network has its own independent P2P overlay \cite{polkadotdesign, rapidchain, free2shard}. An independent P2P overlay provides the flexibility to tailor protocols such as node discovery and routing  to the characteristics and needs of the nodes in that sub-network. For instance, Ethereum has one dedicated P2P overlay for consensus nodes and one for execution nodes \cite{ethP2P}, and heterogeneous parachains in Polkadot each have their own separate P2P overlays \cite{polkadotdesign}. The primary difference between the two architectures is scale: each Ethereum node participates in both overlays, whereas Polkadot nodes each participate in one parachain. In essence, Ethereum's architecture resembles two parallel overlays both spanning the entire network, while Polkadot's P2P network consists of smaller federated overlays dedicated to each parachain.

While these architectures can provide flexibility to accommodate node heterogeneity, they carry new challenges. Smaller P2P overlays can generally disseminte information more efficiently, so existing multi-overlay architectures in blockchain systems facilitate faster information dissemination compared to a single large overlay~\cite{wang2019sok}. However, federated multi-overlay architectures in blockchains often compromise security. Smaller independent P2P overlays are susceptible to network layer attacks, as adversaries can compromise a significant portion of the network with fewer resources \cite{sit2002security}, thereby undermining the overall security of the blockchain system.

\textbf{Contributions.} In this paper, we propose a novel P2P overlay architecture for blockchain systems: \textit{P2P Overlay Domains with Sovereignty (PODS)}, which has the benefits of both overlay architectures.
The overlay is segmented into  \textit{sovereign domains}, each operating independently with its own internal structure and protocols including routing and node discovery.
These protocols can be tailored to the characteristics and needs of the nodes in the domain.
For example, a domain primarily consisting of light clients, experiencing high churn rates, can leverage a random graph topology, benefiting from the resilience of unstructured overlays to node churn~\cite{antwi2022survey}.
Conversely, a domain characterized by low churn but demanding faster information retrieval (e.g., a domain of validator nodes) can adopt a structured approach, prioritizing efficient routing mechanisms \cite{naik2020next}, despite potential longer recovery times during churn.
Crucially, nodes can be part of multiple domains simultaneously.

In particular, PODS overlays can use this reputation to detect domains under attack.
We propose an \textit{active response mechanism}, in which honest nodes join domains under attack, to restore functionality.
Our response mechanism is built on a domain-based reputation system, which PODS naturally facilitates.
Domain-based reputation can offer greater reliability compared to individual-based reputation models, especially in a P2P system when there is churn \cite{josang2007survey}.
With our active response system, a single overlay with multiple domains can even allow each domain to tolerate more adversarial nodes than it could if it was operating as an independent small P2P overlay

To demonstrate the effectiveness of our architecture, we designed and implemented a novel node discovery protocol for PODS called \textit{SovKad}.
SovKad is based on Kademlia~\cite{kademlia}, a widely adopted state-of-the-art single overlay protocol.
To further evaluate the performance of SovKad, we also designed and implemented \textit{FedKad}, a node discovery protocol for a federated multi-overlay architecture, similar to designs proposed for systems like PolkaDot~\cite{polkadotdesign}.
SovKad and FedKad both perform intra-domain lookups as if domains were fully independent.
The key difference is that while FedKad relies on a special \textit{gateway} domain to route inter-domain protcols, SovKad allows each node to contact other domains directly, which has several performance and security advantages. 
We compare the security and performance of SovKad, Kademlia and FedKad. 

We evaluate and compare the security and performance of SovKad, Kademlia and FedKad, using a custom simulator based on PeerSim~\cite{peersim}. 
Our performance measurements show that SovKad outperforms node discovery in a single overlay architecture (Kademlia) and in a multi-overlay architecture (FedKad), providing evidence of the potential of PODS architecture for blockchain systems. 
Our security measurements reveal some of the vulnerabilities of Federated multi-overlay architectures (FedKad), and show that our active response mechanism (in SovKad) substantially improves attack resilience.  
\section{Related Work}
\label{sec:relatedwork}



P2P overlays are essential for secure and efficient communication between nodes in a blockchain network.
Slow information dissemination can be a symptom of a P2P overlay's failure to account for node heterogeneity~\cite{srivatsa2006large}.
As a result, new P2P overlays have been developed to improve performance and provide greater flexibility to accommodate node heterogeneity.

Ethereum has one dedicated P2P overlay for execution and one for consensus client~\cite{ethP2P}.
However, it requires that nodes run both clients simultaneously, so each each P2P overlay is the size of the whole network.
Ethereum trades performance for flexibility and security since this overlay can be huge: its architecture resembles having a single overlay for the whole network.

Sharding blockchains like Polkadot have a separate P2P overlay for each heterogeneous parachain~\cite{polkadotdesign}.
The current architecture allows different overlay topologies and networking protocols in each.
This \textit{Federated P2P Overlay} architecture accommodates node heterogeneity.
Because each subnetwork is smaller, and nodes typically communicate with other nodes in the same sub-network, Federated P2P Overlays offer the performance of a small P2P network.
However, this architecture sacrifices security. 
Each overlay must be secured independently, and smaller P2P overlays are more vulnerable against network layer attacks because an adversary can more easily compromise a significant portion of the network with fewer resources~\cite{sit2002security}.
To ensure the security of the entire blockchain system, we need a unified security system that can provide end-to-end protection across different overlays.

Lin, Ta{\"\i}ani, and Blair explore the advantages of a single system using heterogeneous fully independent P2P overlays~\cite{lin2009exploiting}. 
However, they do not consider the advantages of combining sovereign sub-overlays into a larger PODS-style overlay, including improved attack resistance (see~\cref{sec:attackresponse}).


\section{System Model}
\label{sec:systemmodel}

\subsection{Overlay Model}
We model a P2P overlay as a connected directed graph, denoted $G = \left\langle V,L,D\right\rangle$. Here, \textit{V} is the set of participants, and any \( v_i \in  V \) represents a node in the network. There exists a link \( \left\langle i,j\right\rangle \in  L \) if and only if node \(v_i\) has node \( v_j\) in its routing table.  Furthermore, we assume that there exists a mechanism that assigns each node to one or more domains \( d_i \in D \). The specifics of this mechanism is out of scope of this work. Since participation in multiple domains can be resource-intensive, we upper bound the fraction of nodes in multiple domains (\(p\)) and the maximum number of domains per node (\(m\)). Furthermore, nodes can join and leave the P2P overlay as they please, and the overall churn rate is \(c\). We assume each node has a unique identifier (such as a public key) used by peers to locate one another. 

\subsection{Network Model}
In general, PODS requires no network assumptions beyond those required by the protocols used in each domain. 
However, our node discovery protocol for PODS overlays, SovKad, inherits Kademlia's network requirements~\cite{baumgart2007s}: it requires point-to-point partially synchronous communication.
Any source node can, given a target node's \textit{IP-address}, send a message to the target, and if source and target are honest, it eventually arrives.
After some unknown global synchronization time, messages always arrive within some unknown latency \(\delta\), so SovKad can use increasing timeouts to ensure it eventually finds a latency bound.

Furthermore, we propose an active response protocol, which allows small domains to benefit from the collective failure tolerance of a larger pool of nodes that do not participate in the domain most of the time. 
Our active response protocol requires a synchronous network assumption: messages send from an honest node to an honest node arrive within some known (possibly large) time bound $\Delta$. 
Smaller $\Delta$ assumptions allow faster recovery in the event of an attack, at the expense of false positives (unnecessary attack responses). 
Note that this assumption is not required for SovKad or individual PODS domains: only the attack response mechanism needs to wait for a $\Delta$ timeout.

\subsection{Threat Model}
We assume that the P2P overlay consists of honest and malicious nodes. The fractions of malicious users in the entire network is \textit{f}, meaning that the fraction of honest nodes is $h = 1 - f$. Similar to most blockchain systems \cite{ethP2P, nakamoto2008bitcoin, free2shard, rapidchain}, we assume that \( f < \frac{1}{3} \). While there is no theoretical bound that a P2P overlay cannot withstand a higher fraction of malicious users, there is no need to consider such a fraction if the blockchain on top of it cannot work.

Malicious nodes adopt various strategies to disrupt the activities in the network. We assume that the adversary positions itself within the system (\textit{internal}) and that each colluder node observes part of the system (\textit{local}): the adversary learns messages sent to or from colluder nodes, and knows the overlay topology.
The  adversary is \textit{adaptive} in that it changes its behavior from time to time based on its observations, but it controls some fixed set of colluder nodes.
However, the adversary's computational power is polynomially bounded in the security parameter of the encryption scheme. 

Because our P2P overlay is split in multiple smaller domains, we assume that the attacker's goal is to decrease the activities in the domains. We can model the influence and power of the adversary in each domain as the fraction of colluding nodes in a domain. The quality of a domain is determined by the \textit{Quality-of-the-service (QoS)} settings for the protocols in that domain. The exact strategies of colluding nodes depend on the state of the overlay as we assume an adaptive adversary. For example, the adversary may send a subset (or all) colluding nodes to a single domain and try to corrupt that domain. Finally, we assume that the adversary can also have colluders that co-exist in multiple domains simultaneously subject to \textit{p} and \textit{m}.

\section{P2P Overlay Domains with Sovereignty}
\label{sec:chapter4}

We propose a novel overlay architecture called \textit{P2P Overlay Domains with Sovereignty (PODS)}.
Our approach segments a single overlay into sovereign sub-networks called \textit{domains}.
Each domain operates independently and has its own set of protocols for domain management and communication.
In this section, we first provide an overview of the formal definitions and then discuss the architecture in more depth.



\subsection{Formal Definitions}

\begin{definition}[Node]
Let  $V$ be a set of nodes, $D$ be a set of domains, $L$ be a set of links between two nodes and $u$ be a unique identifier for a node. A node $v \in V$ can be defined as the triplet $\left\langle u, D', L'\right\rangle$ with $D' \subset D$ and $L' \subset L$.  
\end{definition}

\begin{definition}[Link] 
Let $V$ be a set of nodes.
A link $l \in L $ can be defined by tuple $\left\langle v, v'\right\rangle$, where node $v \in V$ has node $v' \in V$ in its routing table.
\end{definition}

\begin{definition}[Protocol]
Let $P$ be a set of protocols, where a protocol $p \in P$ is a set of rules that defines how nodes interact with each other and maintain the overlay.
These rules may include node discovery, link establishment, message routing, and other network operations.
\end{definition}

\begin{definition}[Domain]
Let $V$ be a set of nodes and $P$ be a set of protocols.
A domain $d \in D$ is a tuple $\left\langle V', P'\right\rangle$ with $V' \subset V$ and $P' \subset P$.
\end{definition}

\noindent Based on these definitions, we can define PODS:

\begin{definition}[P2P Overlay Domains with Sovereignty]
Let $V$ be a set of nodes, $L$ be a set of links between those nodes, and $D$ be a set of domains.
Then, a PODS overlay is the triplet $\left\langle V,L,D\right\rangle$. 
\end{definition}

\subsection{Discussion}
The PODS architecture integrates the key advantages of both a single large overlay and a federated multi-overlay architecture.
With a focus on providing flexibility to accommodate node heterogeneity, PODS achieves the performance benefits associated with a small P2P overlay while also maintaining the robust security features typically found in a large P2P overlay.
By combining these strengths, PODS offers a unique approach that optimizes both performance and security within a blockchain system.

\subsubsection{Performance}
Two key factors contribute to PODS' performance advantages relative to a traditional homogeneous overlay.
First, the domains within PODS are designed to be small, enabling efficient information dissemination similar to a smaller P2P overlay.
This improves efficiency compared to a single large P2P overlay~\cite{wang2019sok}. 
Second, domains can optimize protocols for their specific requirements.
For instance, a domain predominantly composed of validator nodes could employ a complete graph structure, ensuring direct connections between all validator nodes without relying on other nodes to relay messages.
This could be valuable for a low-latency consensus protocol.
Conversely, a different domain consisting primarily of light clients could use a random graph topology with reduced connectivity to accommodate the high churn rates typically associated with light clients.
Customized domains further enhance the performance and adaptability of PODS overlays.

The PODS architecture also enables better performance than existing federated multi-overlay architectures in blockchain systems (e.g.,~\cite{kwon2019cosmos, polkadotdesign}). 
This is achieved through two key factors.
First, PODS allows nodes to maintain direct cross-domain connections: there is no need for a single \textit{super-overlay} to relay cross-domain messages.
A super-overlay (e.g., the relay chain in Polkadot~\cite{polkadotdesign}) is  an undesirable point of centralization, and can become a performance and scalability bottleneck. 
Second, PODS allows nodes to participate in multiple domains simultaneously, preventing unnecessary network traffic.
Rather than treating nodes with the same IP address but different overlays as separate entities (resulting in unnecessary network traffic even for self-messaging), PODS allows a single node to be part of multiple domains simultaneously, eliminating the need for such redundant message routing.

\subsubsection{Security}
Multi-overlay architectures in blockchains often compromise security in pursuit of performance.
Smaller independent P2P overlays are susceptible to network layer attacks, as adversaries can compromise a significant portion of the network with fewer resources~\cite{sit2002security}, thereby undermining the overall security of the blockchain system.
Consequently, it becomes easier to perform attacks such as routing, hub or eclipse attacks~\cite{LI2020841, yu2020survey}. 


For protocols to function correctly, P2P overlays require trust mechanisms~\cite{KOUTROULI201247}.
Therefore, trust is a prominent issue in P2P systems, and necessitates the use of trust mechanisms \cite{KOUTROULI201247}. Reputation systems have emerged to address this need.
They constitute an important trust management mechanism in online communities.
In a reputation system, statistics are maintained about the performance of each entity in the system and used to infer how entities are likely to behave in the future \cite{4577789}. 

PODS enables trust mechanisms built on group-based reputation models. 
Specifically, nodes can maintain statistics about domains, rather than individual nodes. 
Group-based reputation models can be more reliable than individual-based models. 
Individual-based models are susceptible to false positives or negatives, as a single malicious or faulty node can significantly impact the overall system's reputation.
In contrast, a group-based model considers the collective behavior of nodes within a domain, providing a more comprehensive assessment of trustworthiness and performance.
This approach also handles the dynamic nature of P2P networks better, as it accounts for the aggregate behavior of nodes over time and reduces the impact of transient nodes \cite{josang2007survey}.
Additionally, a group-based model promotes collaboration and accountability, fostering cooperation among nodes and incentivizing them to maintain a positive reputation collectively~\cite{HENDRIKX2015184}.

Unlike a federated multi-overlay architecture, nodes also maintain links with small sub-set of nodes in other domains, allowing them to build a reputation model of groups of nodes based on interaction with individual nodes.
In a large network, building a reputation model of individual nodes is difficult as it requires significant interaction. Maintaining a reputation of domains allows nodes to make better routing decisions so that information can be propagated more efficiently  and so that attacked domains can be detected and recovered.  
As an example, we develop an active attack response mechanism using a group-based reputation model: when a domain's reputation decays sufficiently, honest nodes step in to restore functionality.

\section{Node Discovery in PODS}
\label{sec:nodediscovery}


To explore the effectiveness of the PODS architecture, we design and implement a \textit{node discovery protocol}: a means by which one node can establish and maintain a link to another~\cite{8456488}. 

\subsection{Motivation}
We focus on node discovery because it is a necessary component for blockchain systems, and it has a strong influence on both performance and security.

The node discovery protocol directly impacts P2P overlay performance~\cite{8456488}.
An efficient node discovery mechanism enables quick and accurate peer lookup, and supports efficient routing.
In a blockchain, it helps nodes quickly identify and establish connections with other nodes responsible for block validation, transaction processing, or information dissemination.
By discovering nearby nodes with high network connectivity and bandwidth, blocks can propagate faster, reducing network latency and improving consensus speed.
Node discovery also impacts transaction dissemination: when a node submits a transaction, it needs to discover peers responsible for transaction validation and inclusion in the blockchain. 

Node discovery also plays a vital role in blockchain security~\cite{8456488}.
It is crucial for maintaining network resilience despite node failures.
As nodes join or leave the network, the network's topology changes, and new nodes need to discover the updated set of participating nodes.
A robust node discovery mechanism ensures that nodes can adapt to these changes and discover alternative peers, enabling uninterrupted information dissemination and maintaining network resilience even in the face of node churn.
Furthermore, attackers can use node discovery protocols to attack the overlay, including as routing attacks and eclipse attacks~\cite{baumgart2007s}. 


We use node discovery to demonstrate a performance advantage of PODS that arises because we account for node heterogeneity. 
Based on data measuring the fraction of transactions across different shards in Ethereum~\cite{das2020efficient}, we estimate that for many domains, intra-domain node lookpus will be much more popular than inter-domain lookups. 
To put it another way, if ``domain $\alpha$'' nodes mostly look up other $\alpha$ nodes, and ``domain $\beta$'' nodes mostly look up other $\beta$ nodes, this makes $\alpha$ and $\beta$ meaningfully heterogeneous in a useful way. 
Our lookup protocols optimize for this: intra-domain lookups are faster than lookups across the whole overlay.

\subsection{Assumptions}
In general, node discovery protocols require that, between any pair of honest nodes, there is a path of existing overlay connections including only honest nodes. 
We also make several assumptions in our node discovery protocols to enable better comparison:
\begin{itemize}
    \item Intra-domain node discovery can function as if the domain were an independent overlay. This requires that between any pair of honest nodes in the same domain, there is a path of existing overlay connections including only honest nodes \textit{within the same domain}.
    \item All domains in PODS architecture operate similar to an independent P2P overlay in federated multi-overlay architecture. For simplicity, we also use  \textit{domain} to refer to an independent overlay in a federated multi-overlay architecture.
    \item All domains in both architectures use Kademlia as their intra-domain node discovery protocol. This is in line with existing literature on sharding-based blockchain systems such as Polkadot~\cite{polkadotdesign} and RapidChain~\cite{rapidchain}. 
    \item The values \textit{p} and \textit{m} are the same in both overlay architectures.
    \item Domains within a federated multi-overlay architecture lack inter-domain attack responses: each domain can only handle attacks using resources and nodes within the domain. This aligns with the findings of a recent study~\cite{mast2022building}, which highlights the absence of recovery mechanisms for domain failures in blockchain systems with a federated multi-overlay architecture.
    
\end{itemize}

\subsection{Goals}
We set out to design a node discovery protocol for PODS that is efficient and resilient. 
We quantify these goals for our experiments in~\cref{sec:metrics}, but intuitively, we want:
\begin{itemize}
\item\textit{Low average routing path length}: minimize the depth of the search used to find nodes.
      This is important for keeping lookup times fast.
\item\textit{Low bandwidth cost}: minimize the number of messages sent per lookup (on average).
      We do not want nodes to refuse to join because the bandwidth cost of participation is too high~\cite{neudecker2015simulation}. 
\item\textit{High resilience}: Node lookups must function quickly and correctly despite network churn and the presence of byzantine nodes. 
In general, protocols can only tolerate a small fraction (such as \(f < \frac13\)) of nodes being byzantine, but crucially, these byzantine nodes may congregate in one domain.
\end{itemize}

\subsection{Preliminaries: Kademlia}
Our node discovery protocol is based on Kademlia~\cite{kademlia}, a well-known DHT protocol used extensively in P2P networks, particularly in file sharing~\cite{ipfs} and content delivery systems~\cite{lavoie2017xor}. 
Recently, blockchain systems have increasingly used Kademlia for node discovery~\cite{polkadotdesign, ethP2P, rapidchain}. 

In Kademlia, each node is assigned a unique $b$-bit identifier (usually $b = 128$ or $160$).
One of the key features of Kademlia is its use of an XOR-based metric to determine the distance between nodes in the network.
The XOR metric has several advantages over other distance metrics, such as Euclidean distance~\cite{kademlia,lavoie2017xor}.

\subsubsection{Routing Tables}
Nodes maintain a routing table that contains the nodeIDs and IP addresses of neighbors in the overlay topology.
Routing tables have more entries for nodeIDs that are closer, based on the XOR metric.
Specifically, each node maintains a routing table consisting of up to $b$ buckets.
Each bucket, known as a \textit{$k$-bucket} in the original paper~\cite{kademlia}, contains up to $k$ entries with relevant information about nodes.
This information depends on the application.
In blockchain systems, it could be the IP address, port number, and timestamp of the last time a node was contacted.
Buckets are arranged as a binary tree, and neighbors are assigned to buckets according to the shortest unique prefix of their nodeIDs. 

\subsubsection{Node Lookups}
In order to begin a lookup, a source node seeking a target node sends a request message to $\alpha$ nodes (usually $\alpha = 3$) from its routing table that are closest to the target (by the XOR metric). 
A node receiving a request responds with $\beta$ nodes from its routing table that are closest to the target (by the XOR metric).
The receiving node also attempts to add the sender to its routing table. 
If the relevant $k$-bucket is full and does not contain the sender, the node sends a ping message to the least recently seen node in the $k$-bucket. 
If this least recently seen node sends a pong message back sufficiently quickly, it becomes the most recently seen node, and the sender is not added to the $k$-bucket. 
Otherwise, the newly discovered node replaces the least recently seen node.

Upon receiving a response message, the source node updates its routing table and sends a new request to the $\beta$ nodes received. 
This process continues until the source node has discovered the $k$ closest nodes to the target node or when it has received the contact information of the target node.
As a side-effect of this lookup operation, nodes that have received a request message also update their routing tables.
In this way, the routing tables of nodes in the network gradually become more complete, allowing efficient and effective message routing.

\subsubsection{Discussion}
One of the key advantages of Kademlia is its efficiency. Because the routing table is organized into buckets based on XOR distance, the number of nodes each bucket represents grows exponentially with the distance from the local node.
This means that the number of nodes queried to find a particular node grows logarithmically with the size of the network~\cite{roos2015determining}.
This is why Kademlia is often referred to as a ``logarithmic'' DHT.

However, as with many structured P2P overlays, Kademlia is vulnerable to overlay attacks.
Two of the most severe overlay attacks on Kademlia are eclipse and routing attacks~\cite{baumgart2007s}:
\begin{itemize}
\item\textit{Eclipse} attacks occur when an attacker effectively cuts off a node from the rest of the network by controlling a sizable number of connections, usually near the target node.
To do this, a Kademlia attacker might control a sizable portion of the nodes that are, in terms of their XOR distance, closest to the target node.
The attacker might alter the target node's routing table, or forge network addresses for other nodes~\cite{baumgart2007s}.
Once the attacker has eclipsed the node, it can prevent it from receiving or sending messages to other nodes in the network, effectively isolating it.
The attacker can use this to disrupt the network or to launch other attacks, such as Sybil or routing attacks.
\item\textit{Routing} attacks occur when an attacker manipulates the routing table of a node.
The attacker causes it to route traffic to malicious nodes instead of legitimate nodes.
In Kademlia, this can be accomplished by sending false or manipulated requests or response messages to a node, causing it to update its routing table with false information.
Once the routing table has been manipulated, the attacker can redirect traffic to nodes under their control, effectively intercepting messages and compromising the security of the network.
\end{itemize}

\subsection{FedKad: Node Discovery in Federated P2P Overlays}
Our Kademlia-inspired node discovery protocol for federated multi-overlay architecture has one super-overlay called the gateway overlay. The gateway overlay contains a sub-set of nodes, called \textit{gateway nodes}, that enable cross-domain communication by relaying messages. This is inspired by Polkadot, where the relay chain forwards messages between parachains~\cite{polkadotdesign}. 

\subsubsection{Routing Tables in FedKad}
In FedKad, non-gateway nodes maintain two separate routing tables: \textit{iDHT} and \textit{xDHT}. Links to nodes that are in the same domain are in the iDHT, and links to gateway nodes are in the xDHT. Gateway nodes maintain a routing table for each domain as they should be able to forward information from one domain to any other. Because gateway nodes are only tasked with forwarding information (i.e., they do not perform node lookups), they can have many routing tables. 

\subsubsection{Node Lookups in FedKad}
In FedKad, there are two types of node lookups: \textit{intra-domain} and \textit{inter-domain} lookups. Intra-domain node lookups are standard Kademlia lookups for a node in the same domain. In inter-domain node lookups, a source node seeks address information of a target node from a different domain. The source node sends a request to a randomly selected gateway node (see \cref{algo:interFedKad}). We do not map gateway nodes to specific domains as this will make it easier for an attacker to make domains unavailable. For example, the attacker can perform DoS attacks on a specific group of gateway nodes, making them unavailable to relay cross-domain messages. 

When a gateway node receives a lookup request from a source node, it updates its routing table according to the Kademlia protocol. If it has the target node in its routing table, it sends the contact information of the target node to the source node. However, if it does not, it forwards the request to the node in its routing table that is closest to the target node (see \cref{algo:interGatewayFedKad}). 

Upon receiving a request message from a gateway node, a node (in the same domain as the target node) initiates a normal Kademlia intra-domain node lookup. When the lookup operation terminates, the final result is propagated back to the gateway node it received the request from, after which the gateway node simply relays the final result back to the source node. 

\subsubsection{Discussion}
FedKad has both advantages and disadvantages in terms of performance and security.

From a performance perspective, average intra-domain routing path length is shorter than in a single P2P overlay. 
Domains are smaller than the whole network, so lookups require fewer hops. 

For inter-domain lookups, FedKad nodes first contact a gateway node. 
This creates additional overhead.
In small overlays, this overhead can reduce performance.
However, for large overlays, this overhead is most likely negligible. 

FedKad has an important performance limitation. 
High churn among gateway nodes can be costly, as nodes in each domain must learn of the new gateway nodes. 
In, for example, Bitcoin~\cite{bitcoinchurn} and Ethereum~\cite{ethP2P}, only a small portion of nodes remain available for over a month, which would make these good candidate gateway nodes. 
In Polkadot, nodes in the relay chain coordinate cross-parachain transactions~\cite{polkadotdesign}, which makes them suitable gateway nodes. 
However, the nodes in the relay chain are occupied with intensive operations, and additional node discovery burden may not be desirable.
Thus, a better approach would be to directly forward lookup requests to nodes from the target domain.

From a security perspective, FedKad's main benefit is to protect honest nodes in one domain against malicious nodes from another domain.
That is, the probability that an intra-domain lookup succeeds in one domain is independent of another domain.

FedKad has also a major security vulnerability.
Attacked domains could remain undetected if malicious nodes perform inter-domain lookups normally. 
For example, the adversary could position all colluder nodes in a single domain and only fail intra-domain lookups.
With all the colluder nodes concentrated in a domain smaller than the whole network, the fraction of colluder nodes in the domain can be much higher than in the network as a whole, so the attacker may exceed Kademlia's failure tolerances. 
For very small domains, or domains with small numbers of honest nodes, it is easy for an attacker to control an arbitrarily large fraction of the domain.

\subsection{SovKad: Node Discovery in PODS}
SovKad is a Kademlia-inspired node discovery protocol for our PODS architecture. The core idea of SovKad, compared to FedKad, is to allow nodes to maintain direct links to a small group of nodes from other domains, in addition to links to nodes from their own domain. Not only does this benefit performance as lookup requests do not have to be routed through gateway nodes, it also allows nodes to build a group-based reputation model about other domains and help repair attacked domains. In SovKad, when an attacked domain experiences compromised integrity, honest nodes proactively join the domain with the objective of restoring the proportion of honest nodes within it. By doing so, the domain becomes accessible to both the honest nodes within the attacked domain and the honest nodes from other domains. This collaborative effort ensures the revitalization of the domain by increasing the presence of trustworthy participants, thereby enhancing the overall security and resilience of the attacked domain.

\subsubsection{Routing Tables in SovKad}
SovKad nodes also maintain two types of routing tables. Similar to FedKad, the iDHT is used to maintain links to nodes within the the same domain. Since nodes in FedKad and SovKad have 128 bits, their iDHTs have at most 128 k-buckets. Unlike FedKad, nodes also maintain a xDHT for every domain it does not participate in. In essence, a node in SovKad maintains a total of \textit{|D|} routing tables, corresponding to the number of domains in the PODS architecture. Given that Ethereum's Kademlia implementation\footnote{https://github.com/ethereum/pydevp2p/blob/develop/devp2p/kademlia.py} typically employs routing tables with at least 256 k-buckets, we distribute the 128 k-buckets across $|D-1|$ domains. Assigning two k-buckets to each domain allows nodes to handle an overlay with up to 64 domains, aligning with the 64 shards in Ethereum~\cite{asgaonkar2022scaling}.

\subsubsection{Node Lookups in SovKad}
Similar to FedKad, intra-domain node lookups in SovKad are standard Kademlia node lookups. Inter-domain lookups, on the other hand, are somewhat different in SovKad. Although nodes have direct links to nodes from other domains, they route the request to a node in the target domain instead of performing the lookup themselves. There are two main reasons for this. First, because iDHTs are larger than xDHTs, performing an intra-domain node lookup is generally more efficient. Second, this design helps honest nodes monitor other domains for attacks, so they can help repair if necessary. When a domain is under attack, meaning too many nodes in that domain are malicious, an adversary may try to deceive nodes that perform inter-domain lookups in that domain. For example, the adversary could behave honestly when receiving inter-domain lookup requests, but dishonestly during intra-domain lookups. This makes it difficult for honest nodes to detect an attacked domain and help repair it. 

When a source node wants to lookup a target from a different domain, it forwards the request to $\alpha$ nodes in the target domain (see \cref{algo:initateinterdomainlookupSovKad}). By sending it to different nodes, we increase the likelihood that an honest node receives the request.  This improves our ability to detect when a domain is under attack: a request fails either when all $\alpha$ nodes are corrupt, or when the domain is under attack.
Upon receiving an inter-domain request, a SovKad node starts an intra-domain lookup and sends the result back to the source node.
 
\subsubsection{Active Attack Response in SovKad}
\label{sec:attackresponse}
When the source node receives the result from one of the $\alpha$ within a specified timeout and the result contains the details of the target node, the lookup is considered a success. When the lookup terminates, the node updates the reputation of the target domain. We select the success rate as a measure of reputation because it is a universal metric with a binary outcome. Other metrics (e.g., latency of the lookup) are not suitable, because they are influenced too much by the internal node discovery protocol and topology of a domain. 

If the reputation of a domain, as measured by success rate of recent requests, falls below a specified threshold \textit{t}, the source node takes action by temporarily joining the attacked domain, provided that the participation constraints $p$ and $m$ are not violated. In this process, the node creates new iDHTs for its current domains. The number of k-buckets in these iDHTs depends on the number of domains the node is involved in, following three rules. First, the number of k-buckets must be equal across all iDHTs. Second, the number of k-buckets for every xDHTs is fixed at 2. Finally, the total number of k-buckets must sum up to 256. Additional details on the algorithm can be found in~\cref{algo:initateinterdomainlookupSovKad}.

\subsubsection{Discussion}
SovKad has three major advantages compared to FedKad.
First, SovKad avoids FedKad's reliance on gateway nodes by allowing nodes to establish links to nodes from other domains. Second, by allowing nodes to participate in multiple domains simultaneously, the probability that a lookup is within the same domain increases. Additionally, self-messaging across domains will not require networking.  Third, it allows nodes to build statistical models for domains. These models can be used to detect whether a domain is under attack, and help repair it.

\section{Evaluation}
\label{sec:evaluation}

\subsection{Methodology}

\subsubsection{Scenarios}
First, we evaluate Kademlia, FedKad and SovKad in an ideal setting where all participants are online and follow the described algorithms (happy path).
These results form a baseline for later experiments.

Second, we evaluate our node discovery protocols when there is network churn.
A large number of nodes can join or leave a P2P network at any time, and studies have shown that churn can have a significant effect on a blockchain’s overall performance \cite{medrano2015performance}.
For instance, high churn rates in Bitcoin have been associated with a 135\% increase in block propagation time~\cite{bitcoinchurn}, which can make the blockchain more vulnerable to forking or double-spend attacks~\cite{gervais2016security}.
It is therefore essential that any node discovery protocol detects and removes stale nodes from its routing tables to prevent offline nodes from being selected to forward information.

Third, we evaluate our protocols with malicious nodes. 
Malicious nodes have a significant negative impact on the performance of any P2P overlay \cite{baumgart2007s},
as they deliberately disrupt the network.
The extent of the damage caused by malicious nodes varies depending on their proportion of the network and their position in the overlay~\cite{free2shard}.
In blockchain systems, malicious nodes attempt to slow down the propagation of transactions or blocks, although in most cases their impact is limited by the assumption that the fraction of malicious nodes is less than \(\frac13\).

However, we will show that even with limited malicious node power, adversaries can still damage the network by strategically positioning themselves in the overlay.
For example, in FedKad and SovKad, the adversary can gather in a single domain, reducing the fraction of honest nodes there.
If they dominate the domain, they can decrease the performance of that domain.
This raises important questions about how different fractions and distributions of malicious nodes affect the performance of intra- and inter-domain node lookups, which are not well explored in prior work. 
Our experiments aim to gain a better understanding of how different fractions and distributions of malicious nodes affect the performance of FedKad and SovKad, and how effective SovKad’s detection and reaction mechanism is.

\subsubsection{Metrics}
\label{sec:metrics}
Here we discuss metrics that allow us to compare whether FedKad and SovKad have met their requirements.
\begin{itemize}
\item \textit{Success Rate}:
The \textit{success rate} is the fraction of lookups that succeed.
We consider a lookup successful if the source node receives the target's contact information within a conservative time bound. 
Low success rate typically means that source nodes cannot find the target nodes, because at least one node along the path towards the target node is unreliable.
A high failure rate may indicate that a node is surrounded by unreliable nodes and that this victim node cannot receive or send information outside its local view.
Therefore, a low success rate indicates that transactions and blocks will not be disseminated effectively.
%
%
\item\textit{Path Length}:
We also monitor the average \textit{number of hops} taken to find each target node.
We do this by calculating the number of nodes that were contacted along the path.
For example, if source node \(v_1\) wants to find target node \(v_3\) and contacts relayer node \(v_2\) which has node \(v_3\) in its routing table, we say this lookup requires 1 hop.
If the source node already has the target node in its routing table, we say that this lookup requires 0 hops.
It is important to keep the average path length low, as this value is directly related to the time it will take for information to reach all nodes.
\item\textit{Bandwidth}:
Finally, we measure the \textit{bandwidth} costs to find a node by calculating the total number of messages that were sent as part of a single lookup procedure.
High bandwidth cost makes it more expensive to participate in the overlay and can also lead to higher latencies or network congestion.
\end{itemize}
\subsubsection{Experimental Approach}
We built a custom open-source simulator to evaluate our overlays. 
We were able to run our simulations on a laptop (an M1 Pro Macbook, with 16GB of RAM), but the hardware used should not affect simulation results.

Our experiments span various numbers of domains, fractions of malicious nodes, and network sizes \(\left|V\right|\) of  1000,  8000, and 16,000.
In line with existing literature on P2P protocols in blockchains~\cite{kadcast}, we simulated 1 hour of network time to allow for cross-study comparison. 
We scale the number of lookups with the size of the network and simulated time. 
On average, each node initiates one lookup per minute. 
We adjust the probability that a search operation is an intra-domain lookup with the number of domains in the overlay. 
We base the fraction of inter-domain node lookups on a recent study that measured the fraction of transactions across different shards in Ethereum~\cite{das2020efficient}: the fraction of inter-domain lookups in our experiments are 5\%, 7.5\%, 10\% and 15\% for 2, 4, 6, and 8 domains,  respectively.
Measurements for our metrics include both intra- and inter- domain lookups.

In our churn experiments, we assume \(\frac14\) of nodes left and \(\frac14\) of nodes joined the network over a time period, a churn rate of \(\frac12\), in line with~\cite{bitcoinchurn}. 
It is hard to provide a specific churn that represents all blockchain systems as there are various factors that influence how long nodes remain online. 
For example, nodes might remain online for a long period if an incentive mechanism is in place for not leaving the network.

We assume \(\frac1{10}\) of nodes are in multiple domains (\textit{gateway nodes}, in FedKad), and that a node can participate in all domains \(\left(m=\left|D\right|\right)\). 
Given that running a node in a blockchain system is resource-intensive, only a small portion
of the network can handle the workload of participating in \(\left|D\right|\) domains.

\subsection{Result and Discussion}

\subsubsection{Results for Happy Path}
The success rate of node lookups in our happy path is 1.0 for all scenarios. Any source node is always able to find any target node for various network sizes and numbers of domains when there is no churn and all nodes are honest. 

\begin{figure}
     \centering
       \captionsetup[subfigure]{justification=centering}
     \begin{subfigure}[b]{0.3\textwidth}
         \centering
         \includegraphics[width=\textwidth]{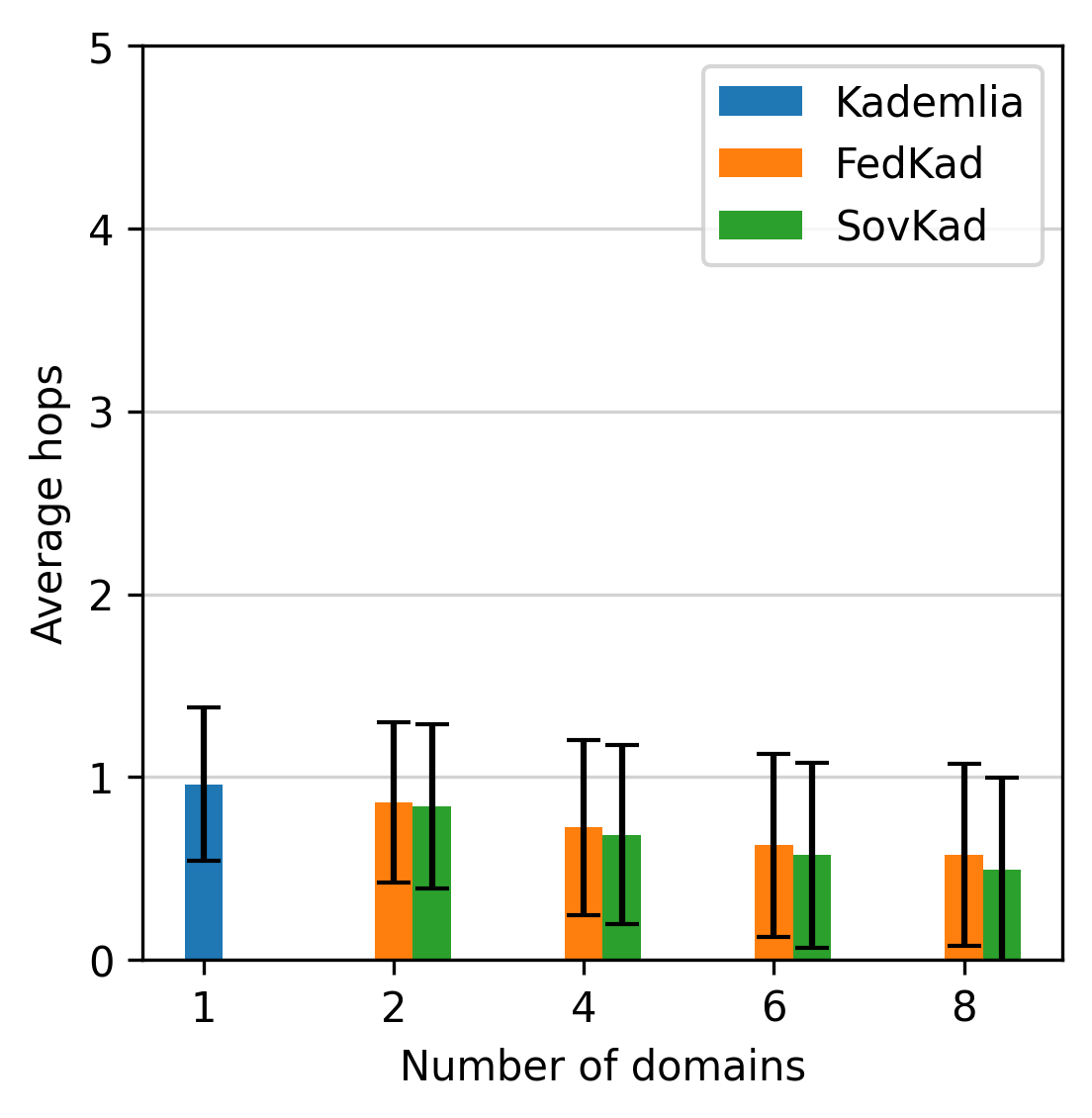}
         \caption{$|V|=1000$}
         \label{fig:happy_hops_1k}
     \end{subfigure}
     \hfill
     \begin{subfigure}[b]{0.3\textwidth}
         \centering
         \includegraphics[width=\textwidth]{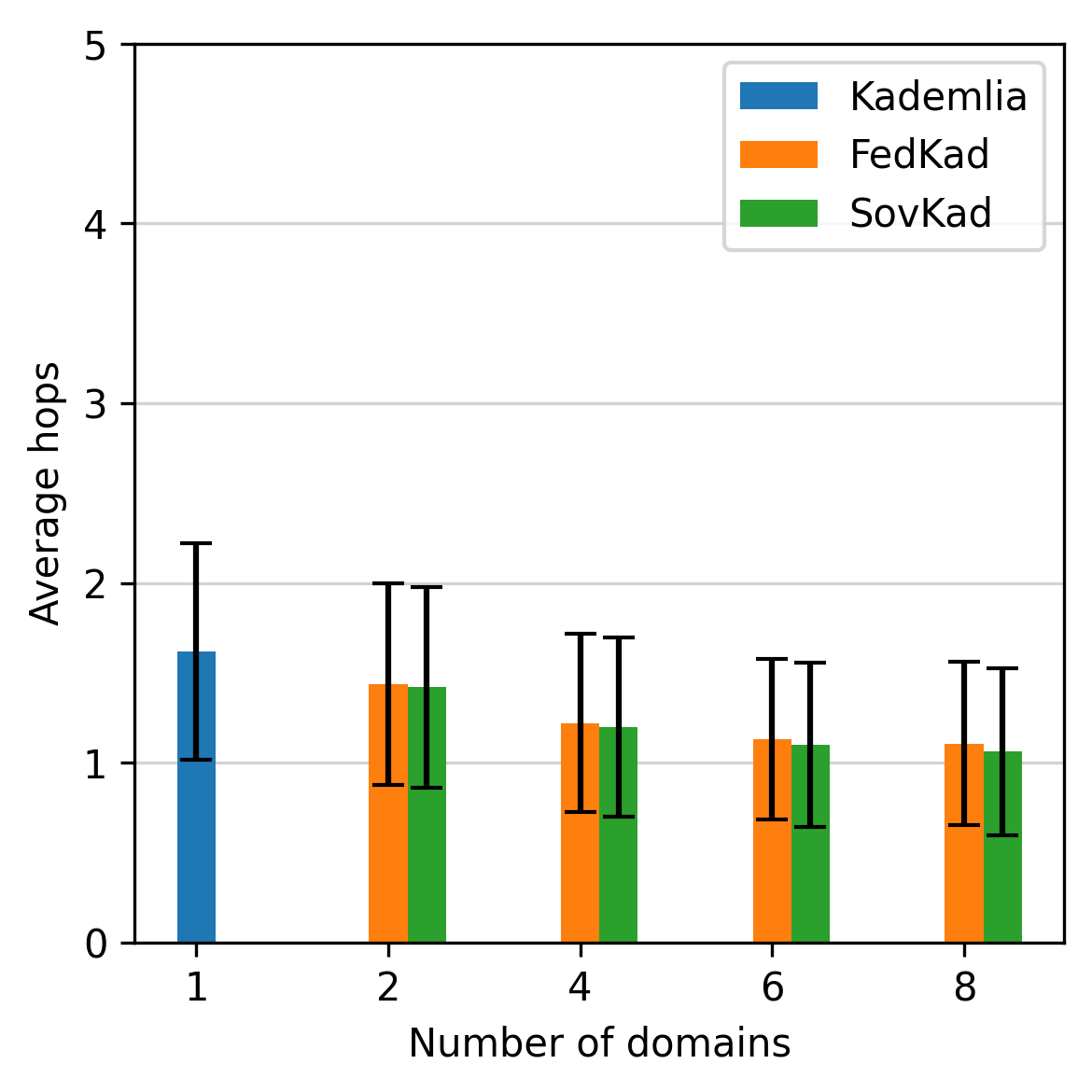}
         \caption{$|V|=8000$}
         \label{fig:happy_hops_8k}
     \end{subfigure}
     \hfill
     \begin{subfigure}[b]{0.3\textwidth}
         \centering
         \includegraphics[width=\textwidth]{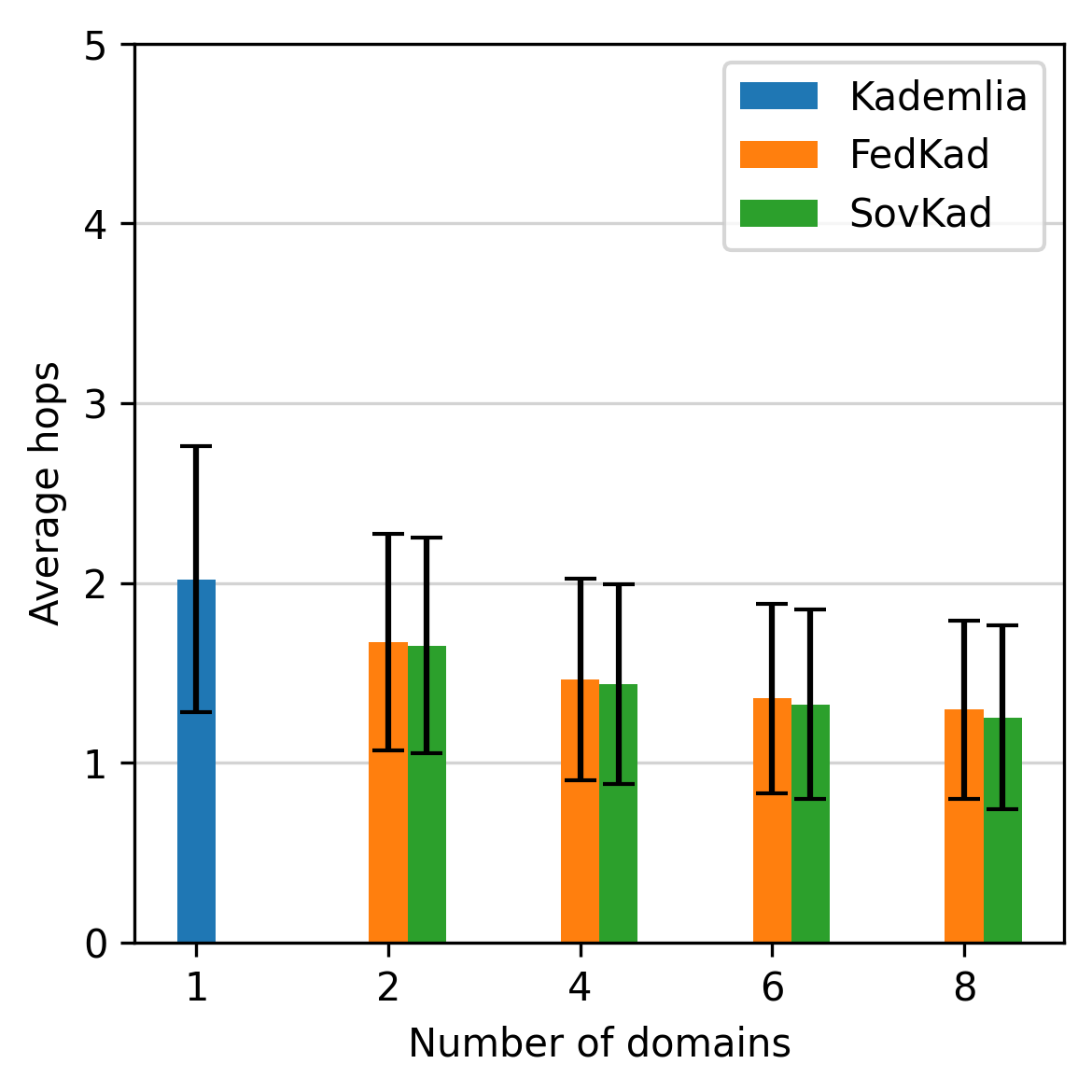}
         \caption{$|V|=16000$}
         \label{fig:happy_hops_16k}
     \end{subfigure}
        \caption{Average routing path length in Kademlia, FedKad and SovKad for happy path. Bars represent standard deviation.  }
        \label{fig:happy_hops}
\end{figure}

Furthermore, FedKad and SovKad outperform Kademlia in terms of  routing path length (see \cref{fig:happy_hops}). The results for FedKad and SovKad are much alike, which is unsurprising since both algorithms are very similar in the happy path. For larger network sizes, the variance in the number hops for FedKad and SovKad is lower than for Kademlia, indicating that the paths in an federated multi-overlay and PODS architecture are more consistent in length,  and that such topologies provide more optimal paths for all nodes rather than for some. 

\begin{figure}
     \centering
       \captionsetup[subfigure]{justification=centering}
     \begin{subfigure}[b]{0.3\textwidth}
         \centering
         \includegraphics[width=\textwidth]{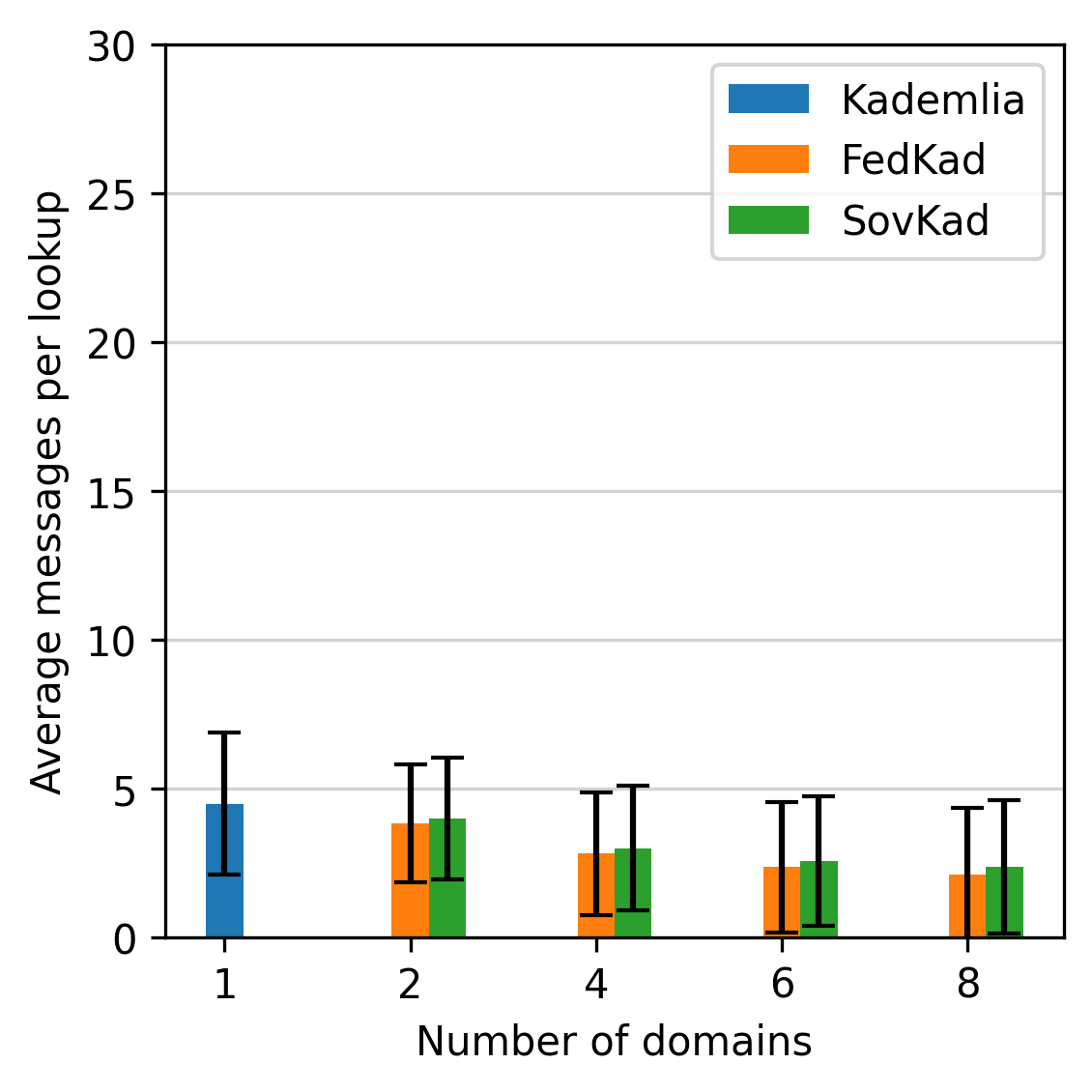}
         \caption{$|V|=1000$}
         \label{fig:happy_messages_1k}
     \end{subfigure}
     \hfill
     \begin{subfigure}[b]{0.3\textwidth}
         \centering
         \includegraphics[width=\textwidth]{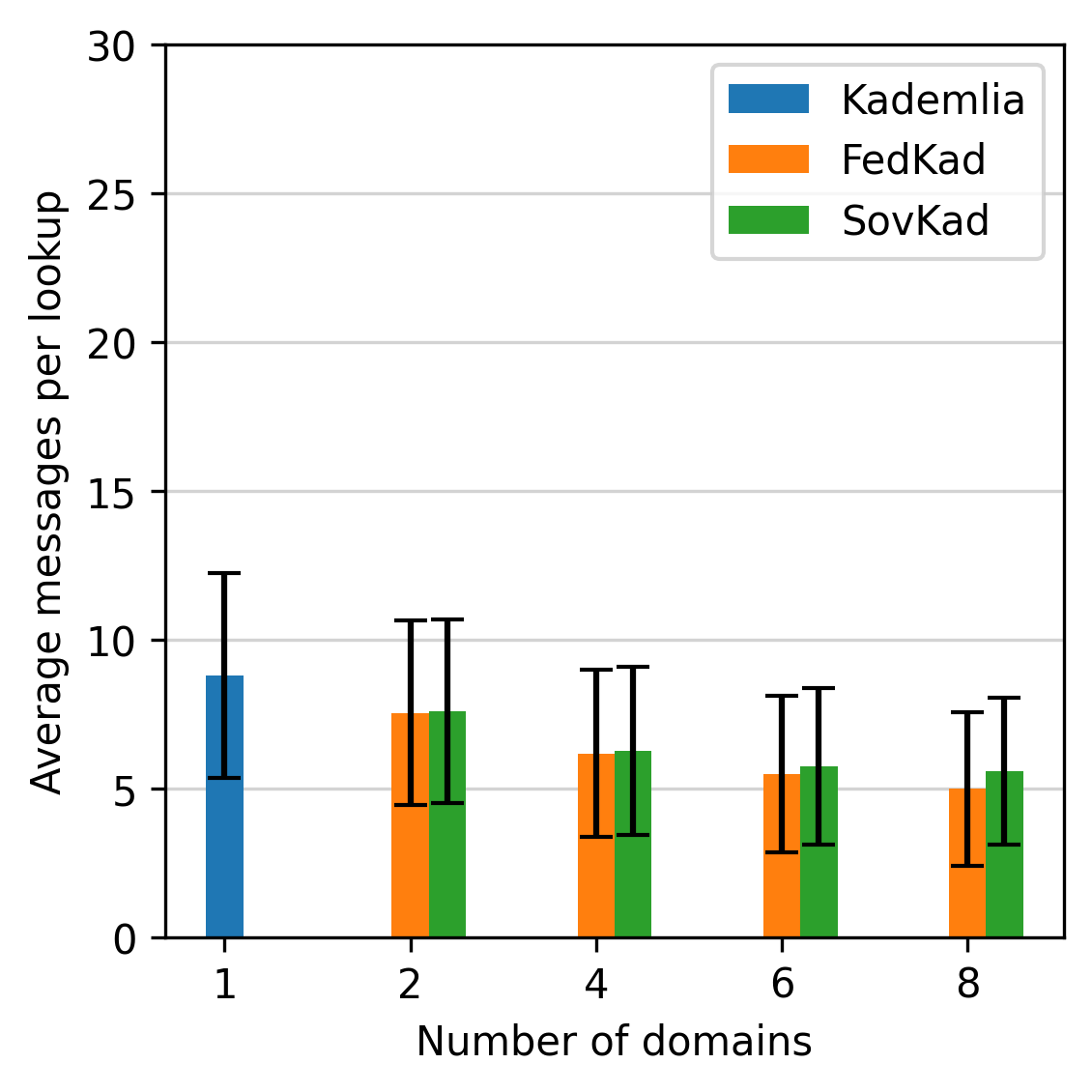}
         \caption{$|V|=8000$}
         \label{fig:happy_messages_8k}
     \end{subfigure}
     \hfill
     \begin{subfigure}[b]{0.3\textwidth}
         \centering
         \includegraphics[width=\textwidth]{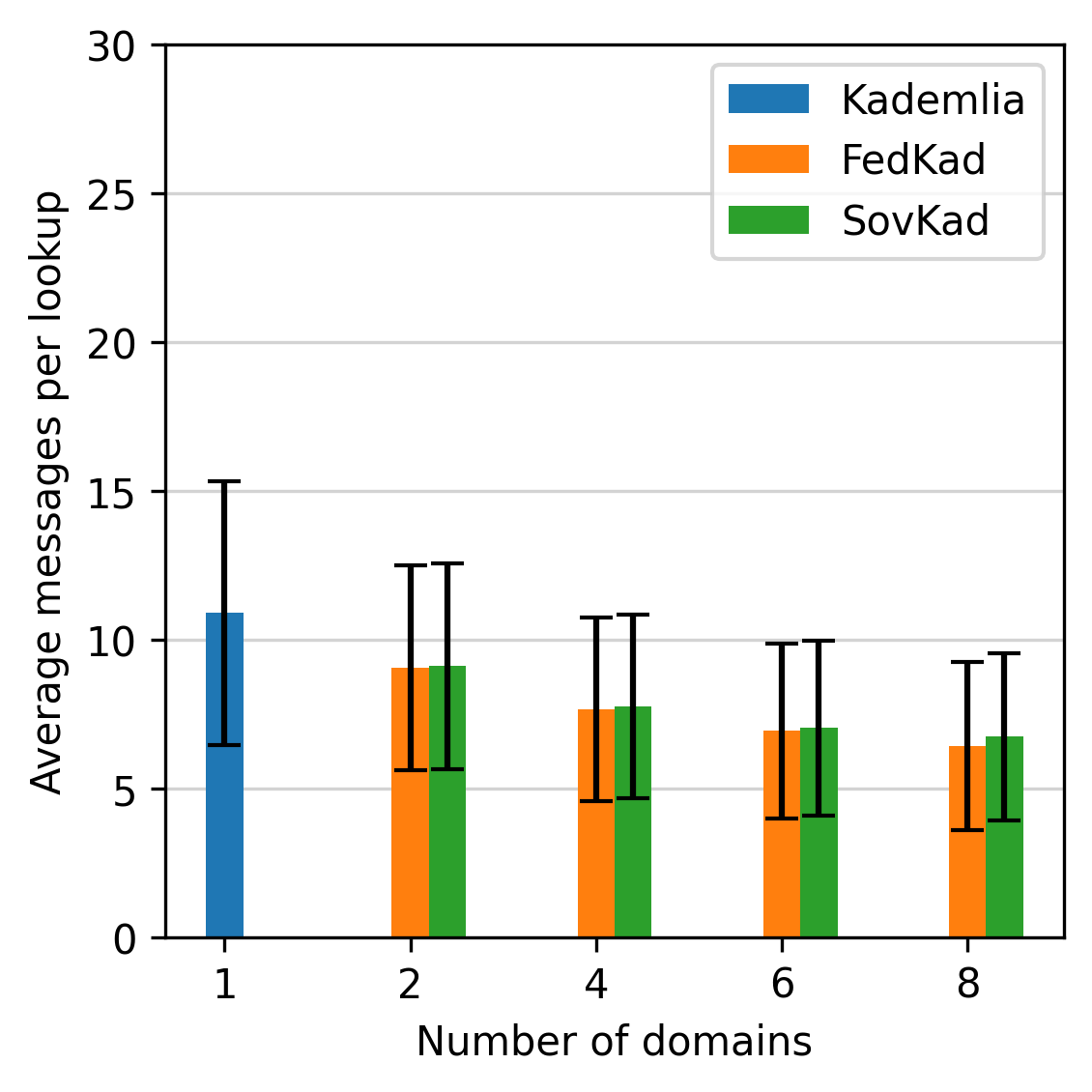}
         \caption{$|V|=16000$}
         \label{fig:happy_messages_16k}
     \end{subfigure}
        \caption{Bandwidth cost (total messages sent per lookup) in Kademlia, FedKad and SovKad for happy path. Bars represent standard deviation. }
        \label{fig:happy_messages}
\end{figure}

\Cref{fig:happy_messages} displays the average bandwidth cost of Kademlia, FedKad and SovKad under different network sizes and different numbers of domains. For all three overlays, larger networks use more bandwidth per lookup. Longer path require contacting more nodes, which uses more bandwidth. Bandwidth variance exhibits a similar pattern to that of the routing path length variance. 

\subsubsection{Results for Churn Scenario}
Our simulations indicate that Kademlia, FedKad and SovKad perform well during network churn. In contrast to earlier findings~\cite{medrano2015performance}, our results suggest that nodes in Kademlia-based protocols are always able to find the target node for a churn rate of $\frac12$. Several factors that may contribute to this mismatch:
\begin{itemize}
    \item \textit{Difference in k-bucket size}: We set bucket-size ($k$) to 20 rather than 8.
    Larger routing tables make nodes more likely to select a candidate that is close to the target node.
    \item \textit{Difference in target node selection}: Our simulator does not simulate target nodes going offline after they are selected for a node lookup.
    It is unclear whether prior work~\cite{medrano2015performance} does the same.
    If not, it could result in lower success rate, as the lookups for offline nodes trivially fail. 
    \item \textit{Difference in retry behavior}: It is unclear whether ~\cite{medrano2015performance} retries another node after a timeout (when the receiver of a request is offline).
    In our simulations, nodes try 3 times before they give up and mark the lookup as unsuccessful. 
\end{itemize}

Furthermore, we found churn made a little difference in routing path length and bandwidth for Kademlia, FedKad and SovKad.
This demonstrates our protocols abilities to adapt to changes in the network, ensuring efficient communication. 

\subsubsection{Results for Byzantine Nodes Scenarios}

\begin{figure}[h]
     \centering
       \captionsetup[subfigure]{justification=centering}
     \begin{subfigure}[b]{0.3\textwidth}
         \centering
         \includegraphics[width=\textwidth]{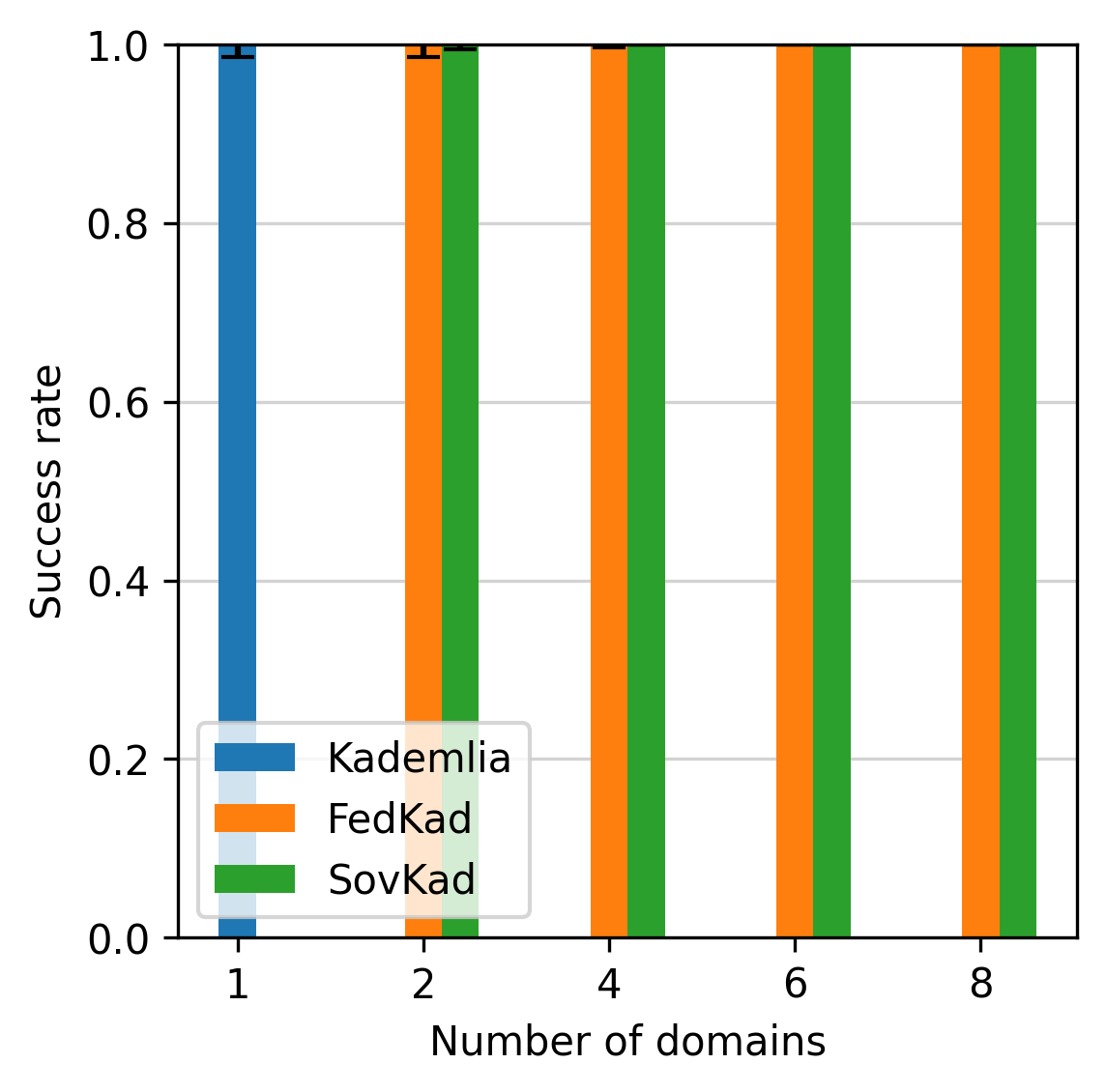}
         \caption{$|V|=1000$}
         \label{fig:byzantine_uniform_succesrate_1k}
     \end{subfigure}
     \hfill
     \begin{subfigure}[b]{0.3\textwidth}
         \centering
         \includegraphics[width=\textwidth]{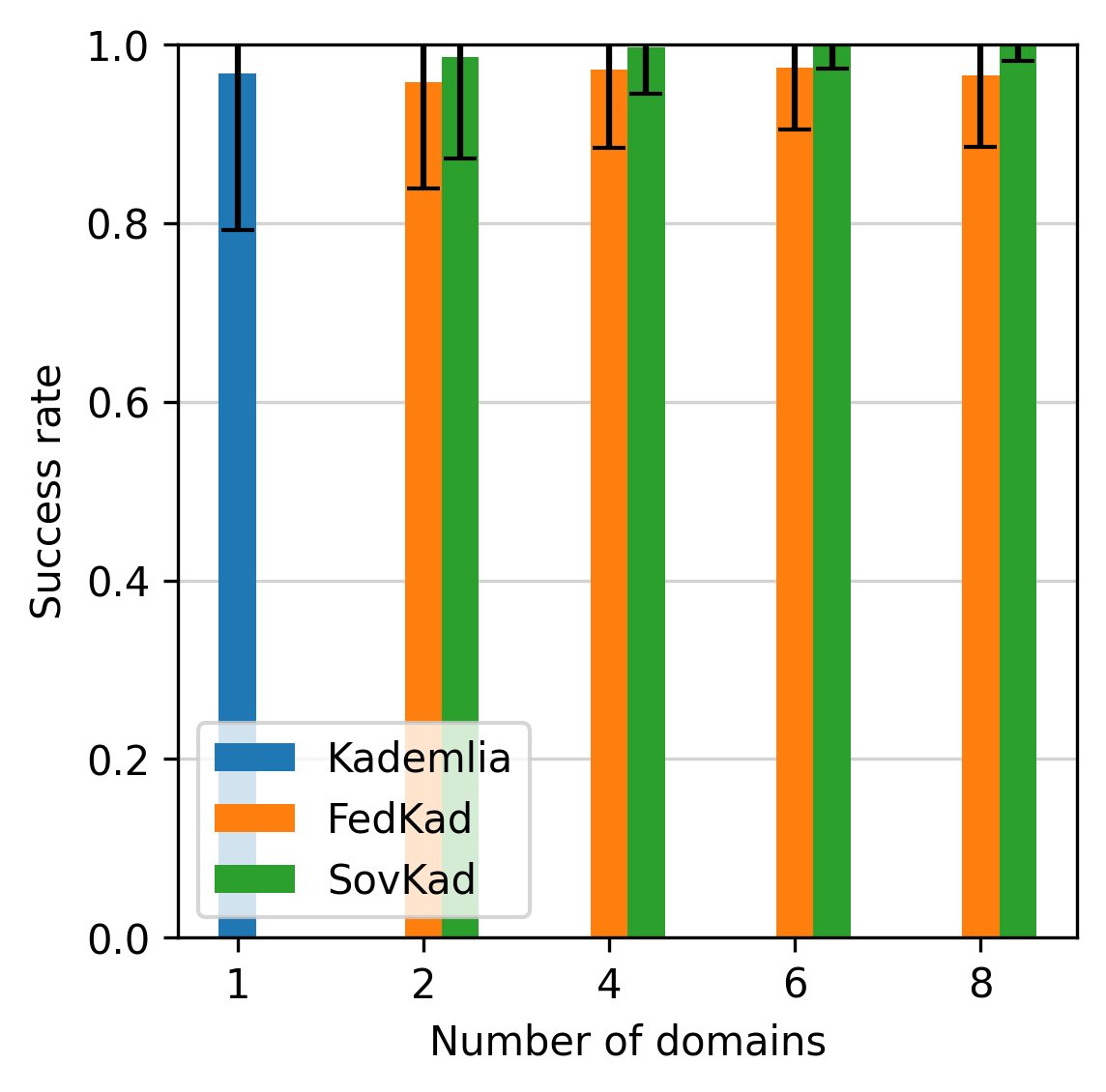}
         \caption{$|V|=8000$}
         \label{fig:byzantine_uniform_succesrate_8k}
     \end{subfigure}
     \hfill
     \begin{subfigure}[b]{0.3\textwidth}
         \centering
         \includegraphics[width=\textwidth]{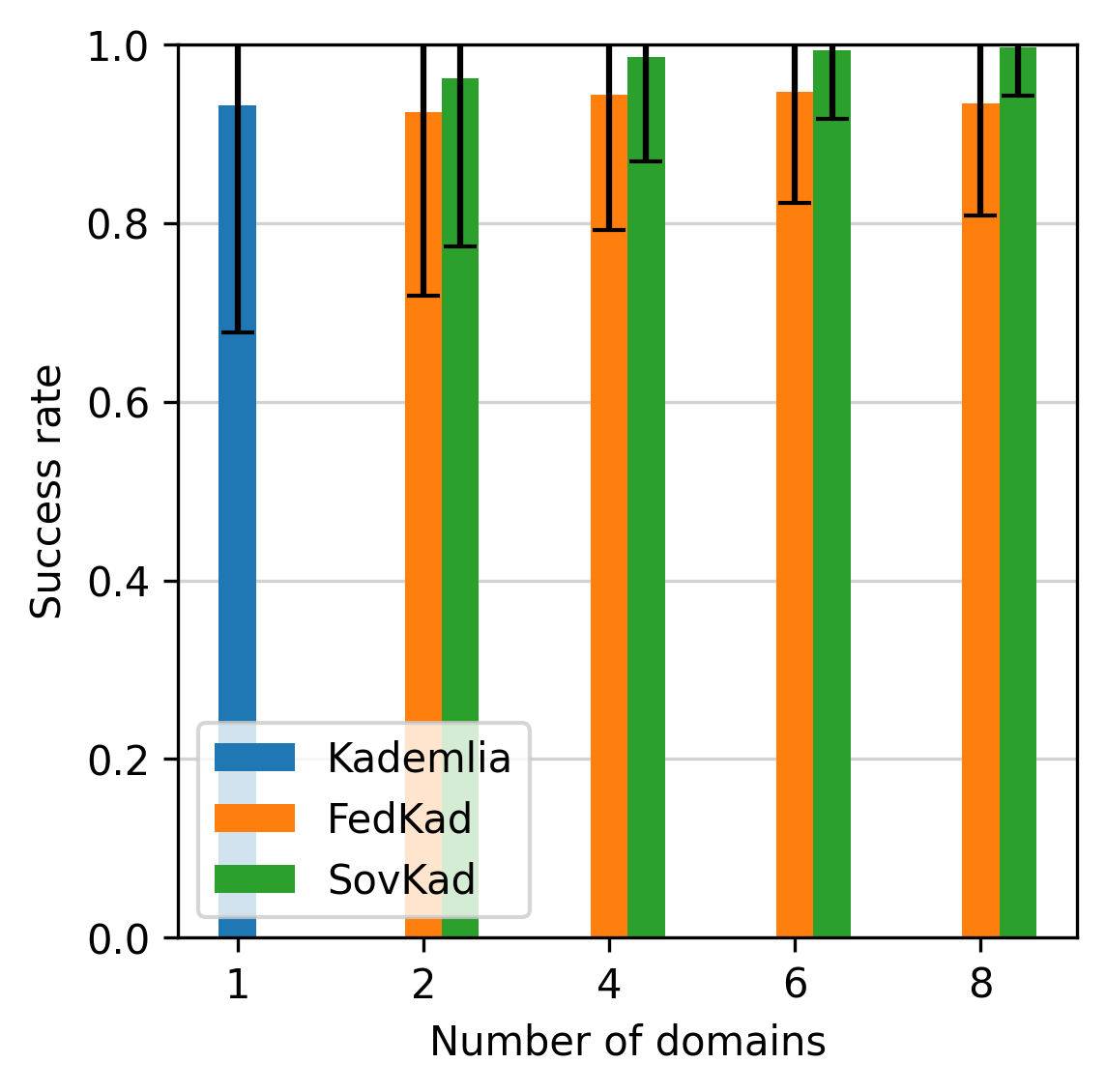}
         \caption{$|V|=16000$}
         \label{fig:byzantine_uniform_succesrate_16k}
     \end{subfigure}
        \caption{Average success rate lookups in a single overlay (Kademlia), a federated multi-overlay (FedKad), and a PODS (SovKad). The analysis is conducted for a scenario with  $f=0.3$, where all colluder nodes are uniformly distributed across the domains. Bars represent standard deviation. }
        \label{fig:byzantine-uniform-success}
\end{figure}

\begin{figure}
     \centering
       \captionsetup[subfigure]{justification=centering}
     \begin{subfigure}[b]{0.3\textwidth}
         \centering
         \includegraphics[width=\textwidth]{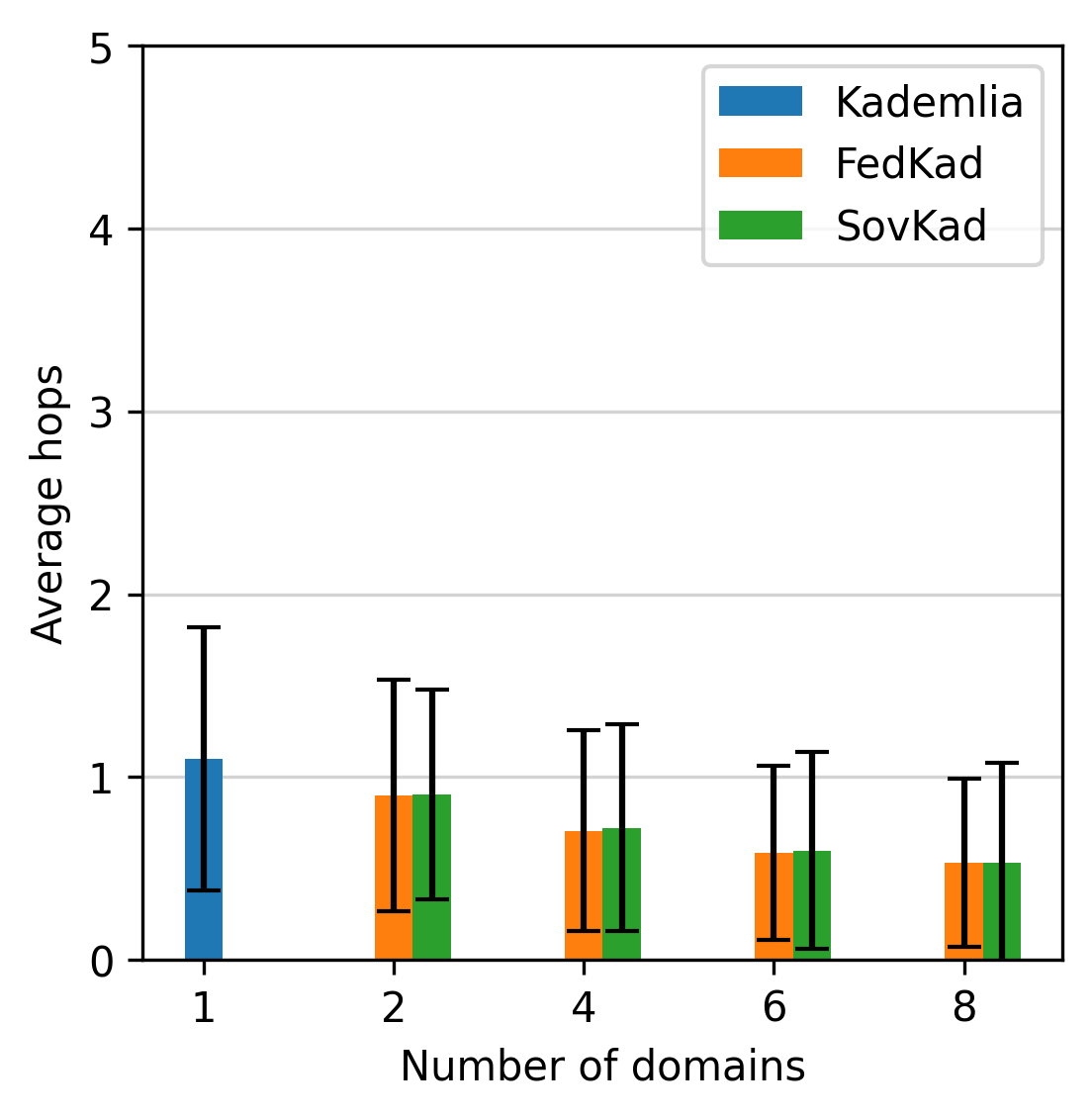}
         \caption{$|V|=1000$}
         \label{fig:byzantine_uniform_hops_1k}
     \end{subfigure}
     \hfill
     \begin{subfigure}[b]{0.3\textwidth}
         \centering
         \includegraphics[width=\textwidth]{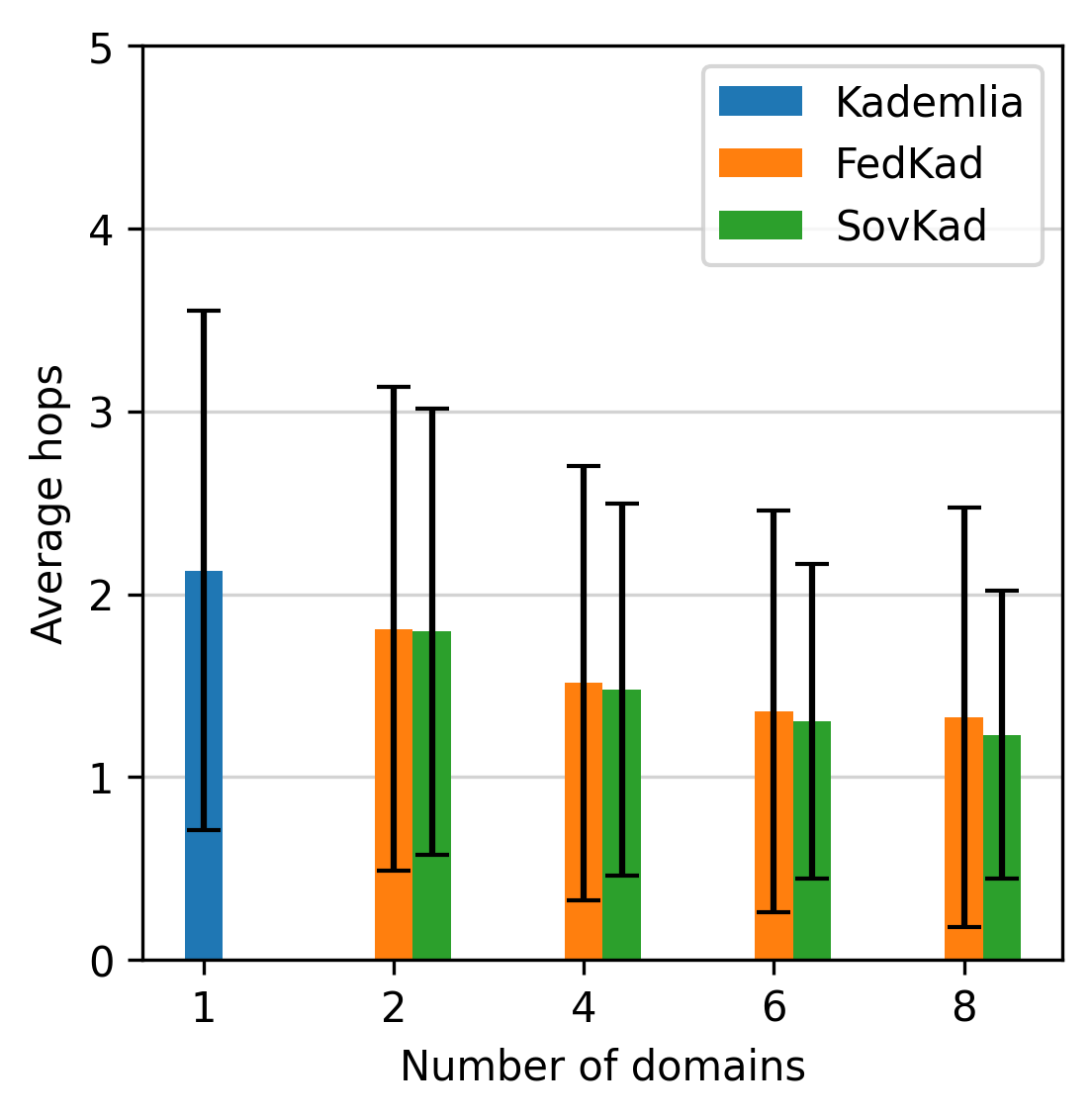}
         \caption{$|V|=8000$}
         \label{fig:byzantine_uniform_hops_8k}
     \end{subfigure}
     \hfill
     \begin{subfigure}[b]{0.3\textwidth}
         \centering
         \includegraphics[width=\textwidth]{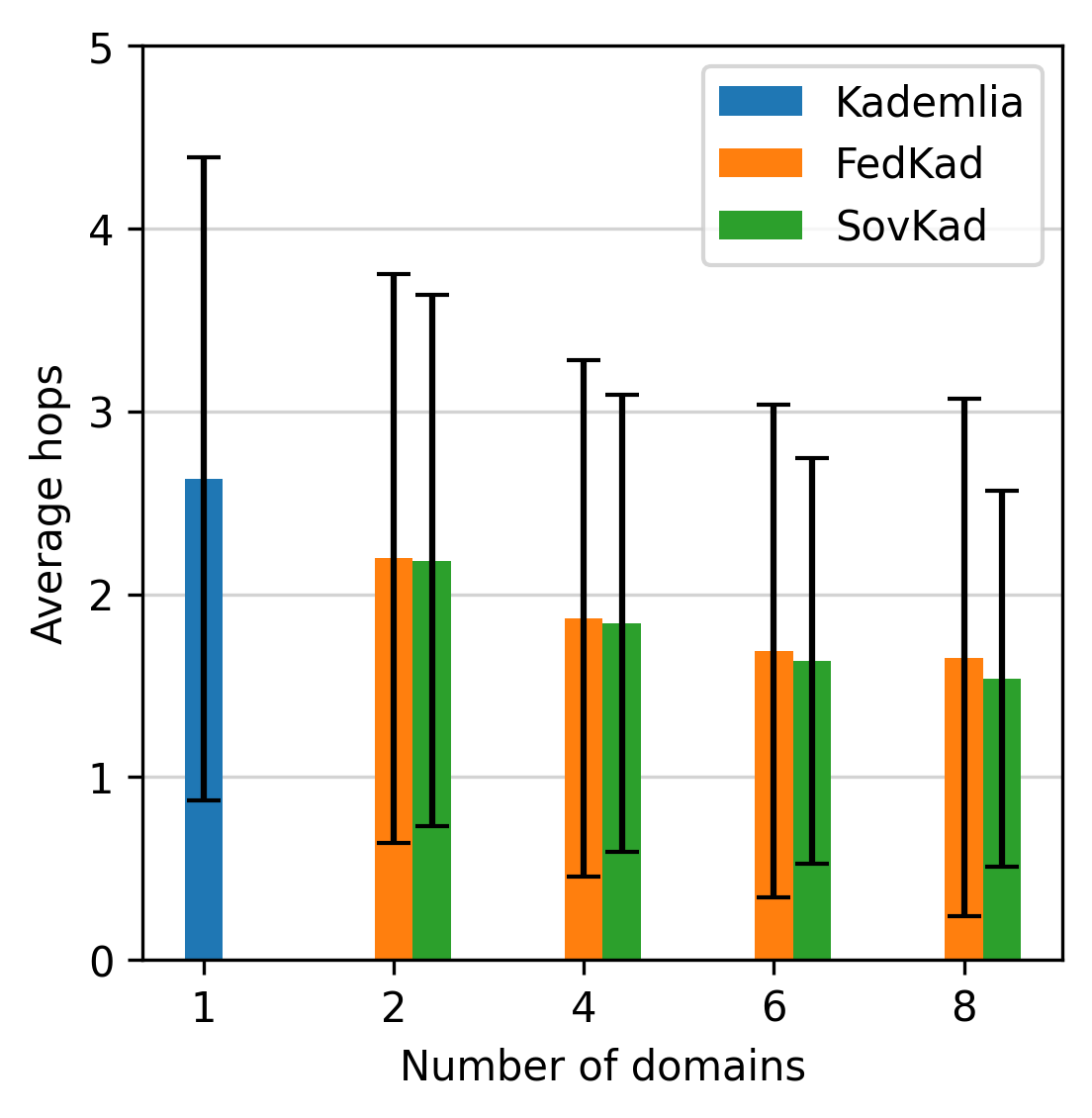}
         \caption{$|V|=16000$}
         \label{fig:byzantine_uniform_hops_16k}
     \end{subfigure}
        \caption{Average routing path length per lookup, measured as the total number hops, in a single overlay (Kademlia), a federated multi-overlay (FedKad), and a PODS (SovKad). The analysis is conducted for a scenario with  $f=0.3$, where all colluder nodes are uniformly distributed across the domains. Bars represent standard deviation. }
        \label{fig:byzantine_uniform_hops}
\end{figure}

\begin{figure}
     \centering
       \captionsetup[subfigure]{justification=centering}
     \begin{subfigure}[b]{0.3\textwidth}
         \centering
         \includegraphics[width=\textwidth]{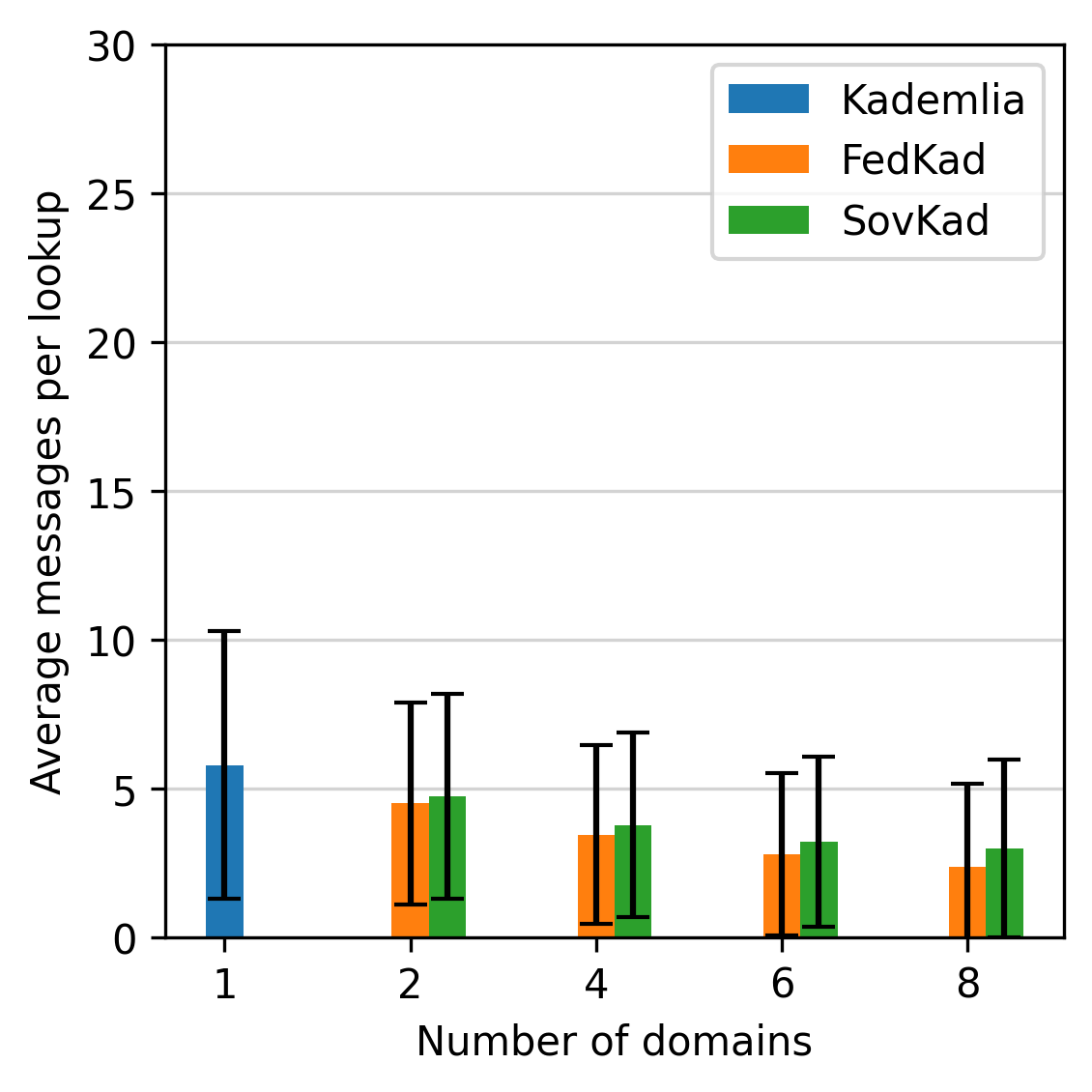}
         \caption{$|V|=1000$}
         \label{fig:byzantine_uniform_messages_1k}
     \end{subfigure}
     \hfill
     \begin{subfigure}[b]{0.3\textwidth}
         \centering
         \includegraphics[width=\textwidth]{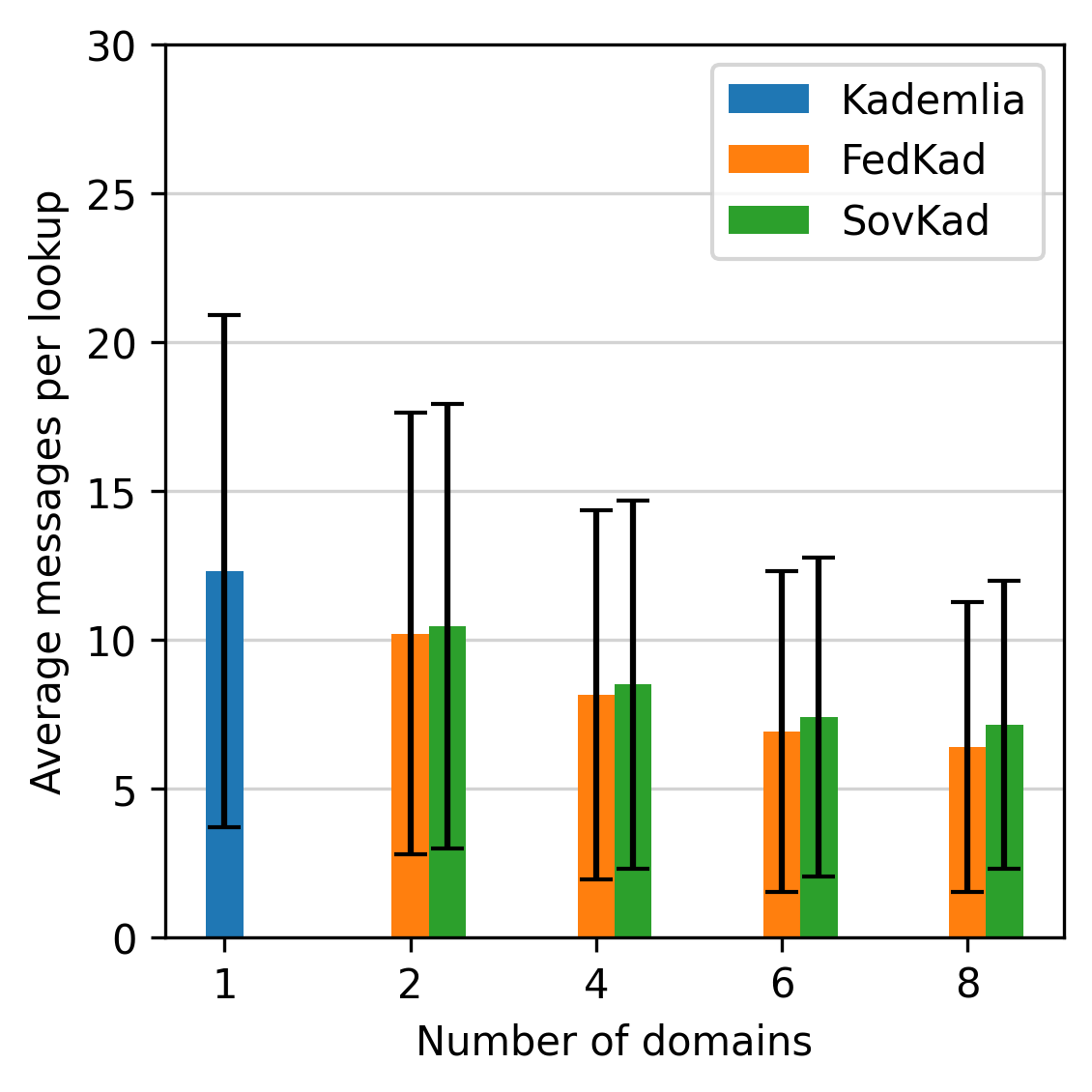}
         \caption{$|V|=8000$}
         \label{fig:byzantine_uniform_messages_8k}
     \end{subfigure}
     \hfill
     \begin{subfigure}[b]{0.3\textwidth}
         \centering
         \includegraphics[width=\textwidth]{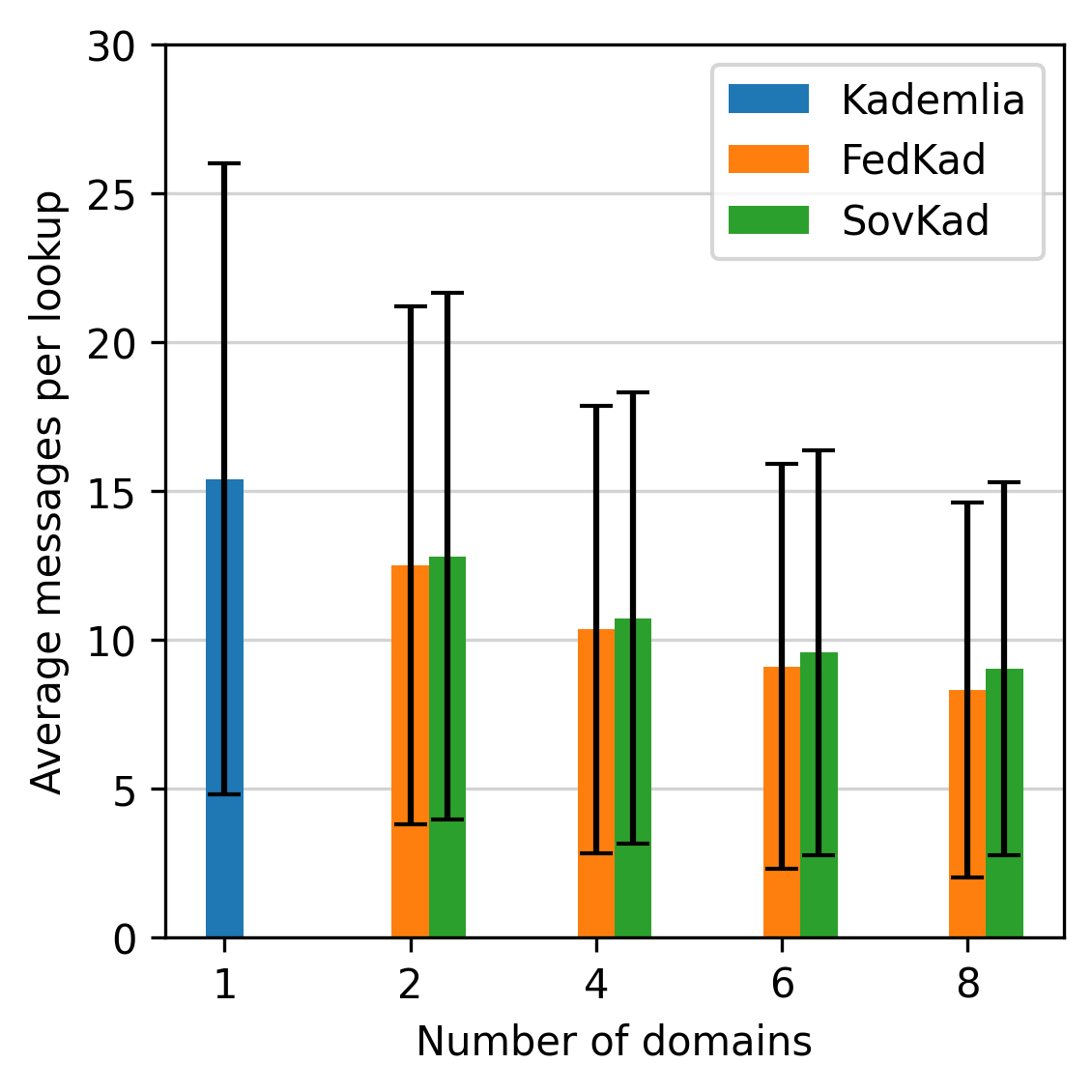}
         \caption{$|V|=16000$}
         \label{fig:byzantine_uniform_messages_16k}
     \end{subfigure}
        \caption{Average bandwidth cost per lookup, measured as the total number of messages sent,  in a single overlay (Kademlia), a federated multi-overlay (FedKad), and a PODS (SovKad). The analysis is conducted for a scenario with  $f=0.3$, where all colluder nodes are uniformly distributed across the domains. Bars represent standard deviation.}
        \label{fig:byzantine-uniform-messages}
\end{figure}

\paragraph{Uniform Distribution}
First, we ran tests in which  30\% of nodes collude with the adversary ($f=0.3$), but these colluder nodes are distributed uniformly at random, and do not change domains. 
Unlike our Happy Path tests, these attacks do manage to make a few lookups fail at larger network sizes (see~\cref{fig:byzantine-uniform-success}).
While FedKad's success rate remained similar to Kademlia's (albeit perhaps with lower variance), SovKad's active attack response mechanism was able to repair domains with an unfortunately high colluder population, and achieve higher success rates. 
Path lengths for successful lookups were generally longer in the attack scenario than the happy path, although they displayed similar relative performance between protocols (see~\cref{fig:byzantine_uniform_hops}). 
Bandwidth cost (including messages sent in failed requests) was likewise higher in the attack scenario, with similar relative performance (see~\cref{fig:byzantine-uniform-messages}).

\begin{figure}
     \centering
       \captionsetup[subfigure]{justification=centering}
     \begin{subfigure}[b]{0.3\textwidth}
         \centering
         \includegraphics[width=\textwidth]{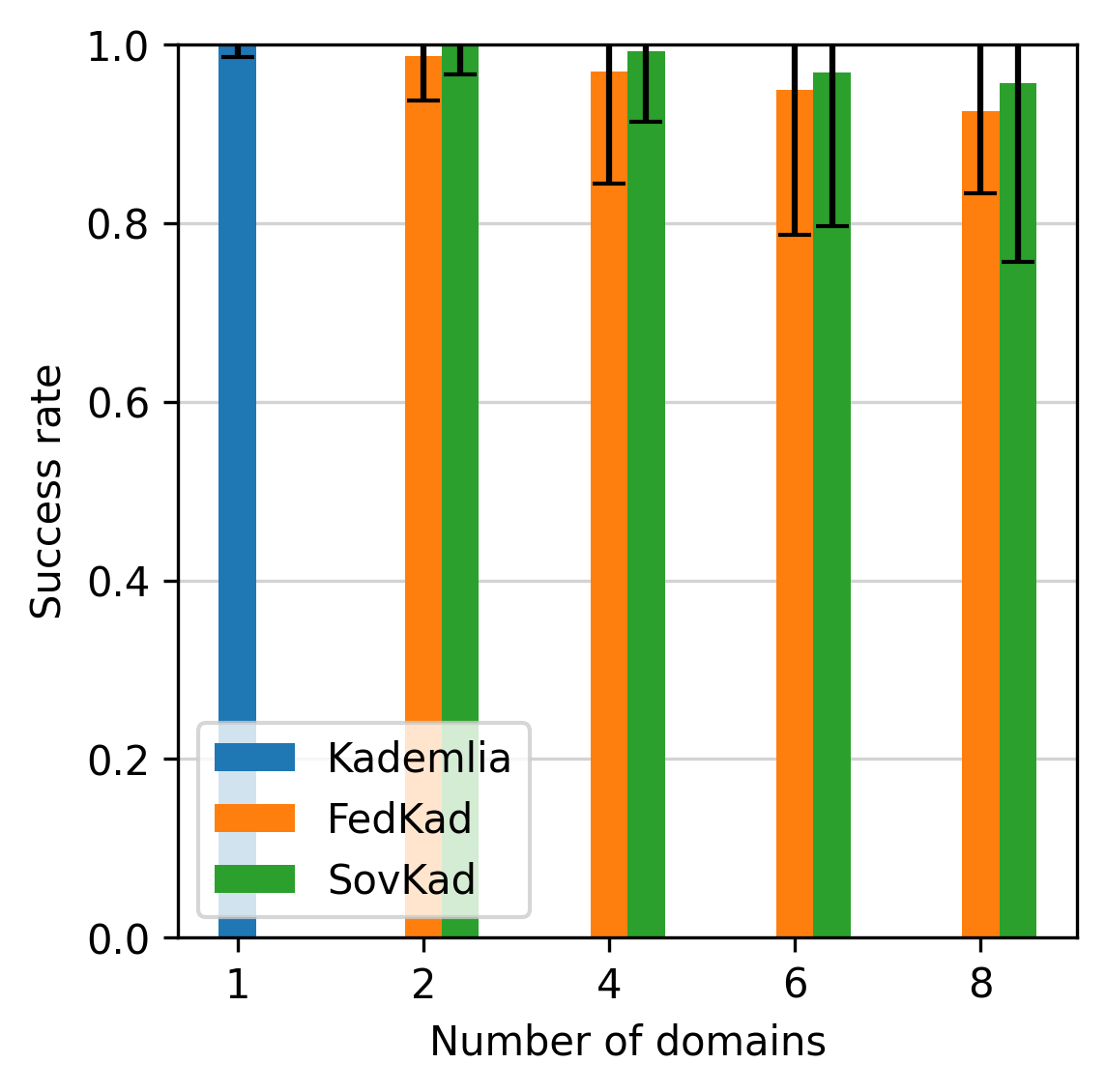}
         \caption{$|V|=1000$}
         \label{fig:byzantine_congregated_succesrate_1k}
     \end{subfigure}
     \hfill
     \begin{subfigure}[b]{0.3\textwidth}
         \centering
         \includegraphics[width=\textwidth]{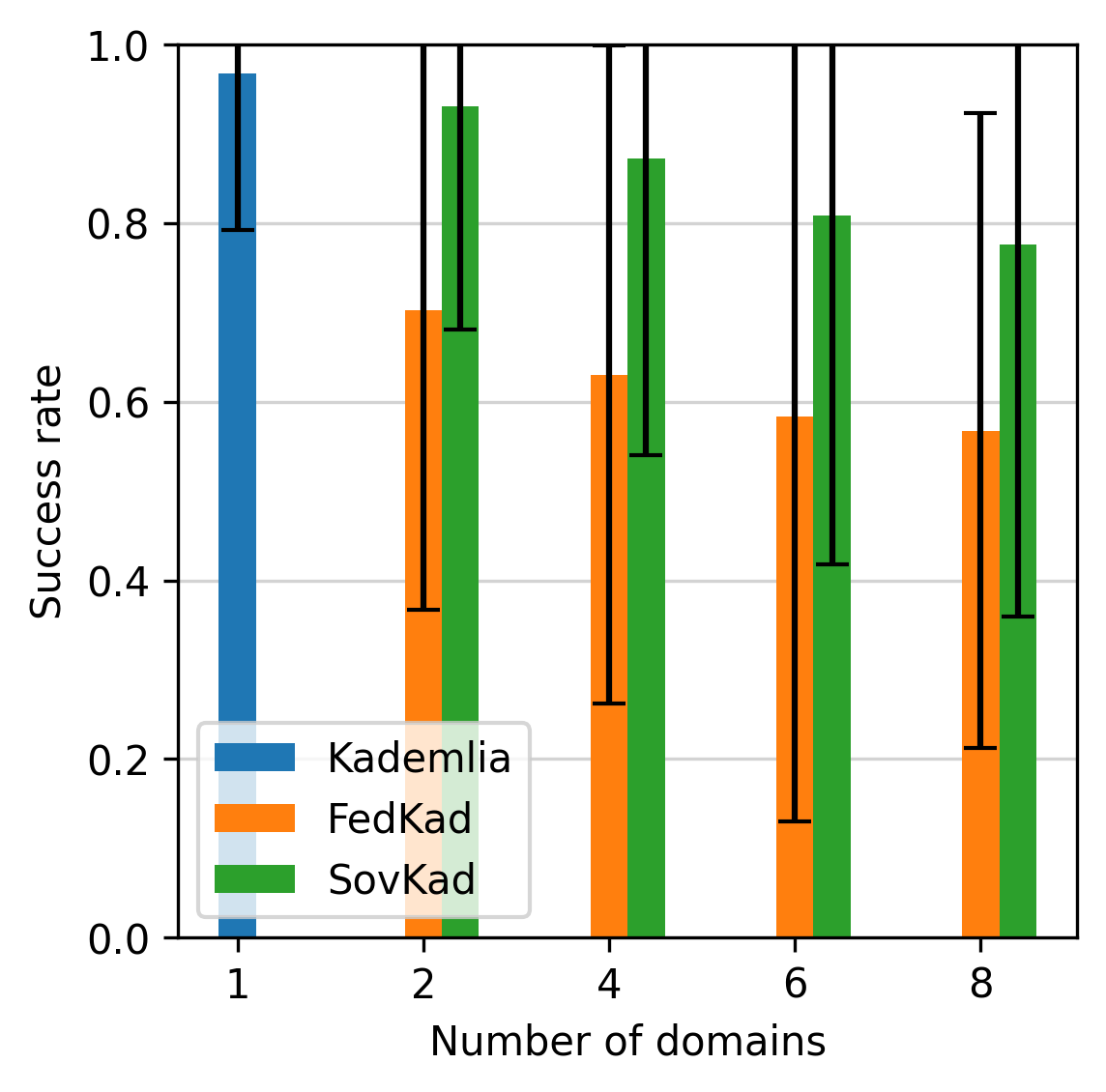}
         \caption{$|V|=8000$}
         \label{fig:byzantine_congregated_succesrate_8k}
     \end{subfigure}
     \hfill
     \begin{subfigure}[b]{0.3\textwidth}
         \centering
         \includegraphics[width=\textwidth]{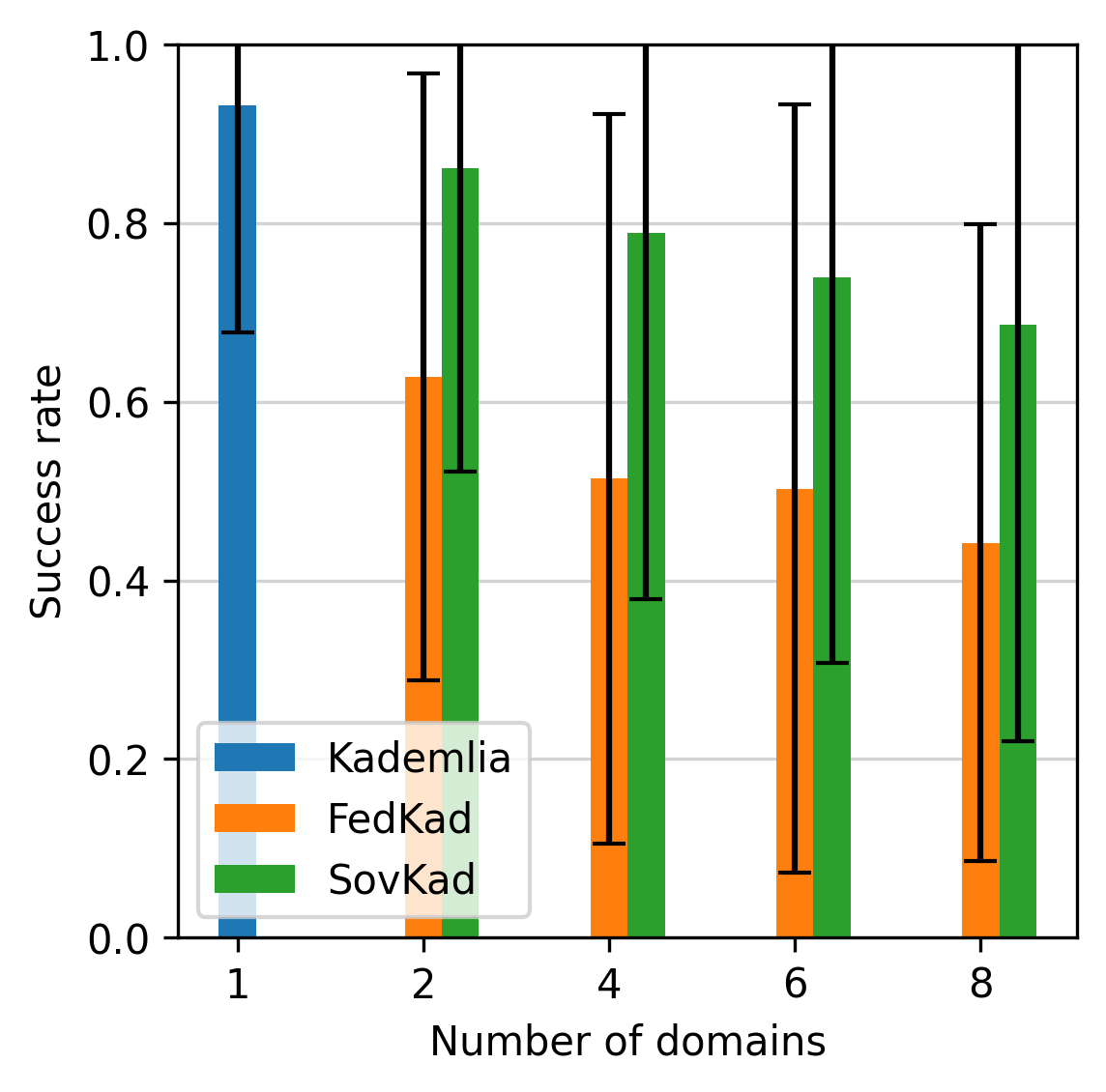}
         \caption{$|V|=16000$}
         \label{fig:byzantine_congregated_succesrate_16k}
     \end{subfigure}
        \caption{Average success rate lookups in a single overlay (Kademlia), the attacked domain in a federated multi-overlay (FedKad), and the attacked domain in a PODS (SovKad). The analysis is conducted for a scenario with  $f=0.3$, where all colluder nodes clustered in one of the domains. Bars represent standard deviation.}
        \label{fig:congregate-success}
\end{figure}

\begin{figure}
     \centering
       \captionsetup[subfigure]{justification=centering}
     \begin{subfigure}[b]{0.3\textwidth}
         \centering
         \includegraphics[width=\textwidth]{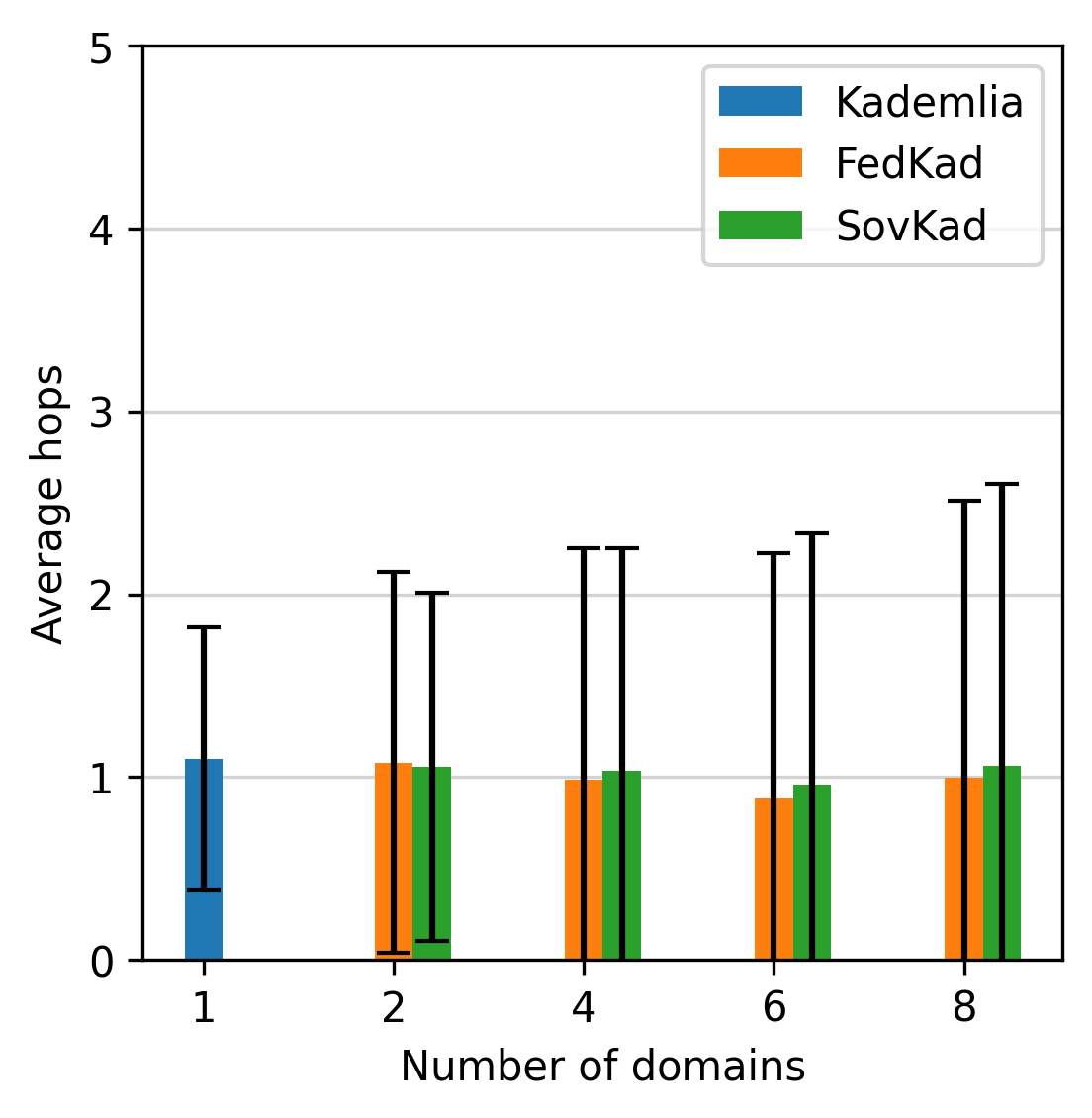}
         \caption{$|V|=1000$}
         \label{fig:byzantine_congregated_hops_1k}
     \end{subfigure}
     \hfill
     \begin{subfigure}[b]{0.3\textwidth}
         \centering
         \includegraphics[width=\textwidth]{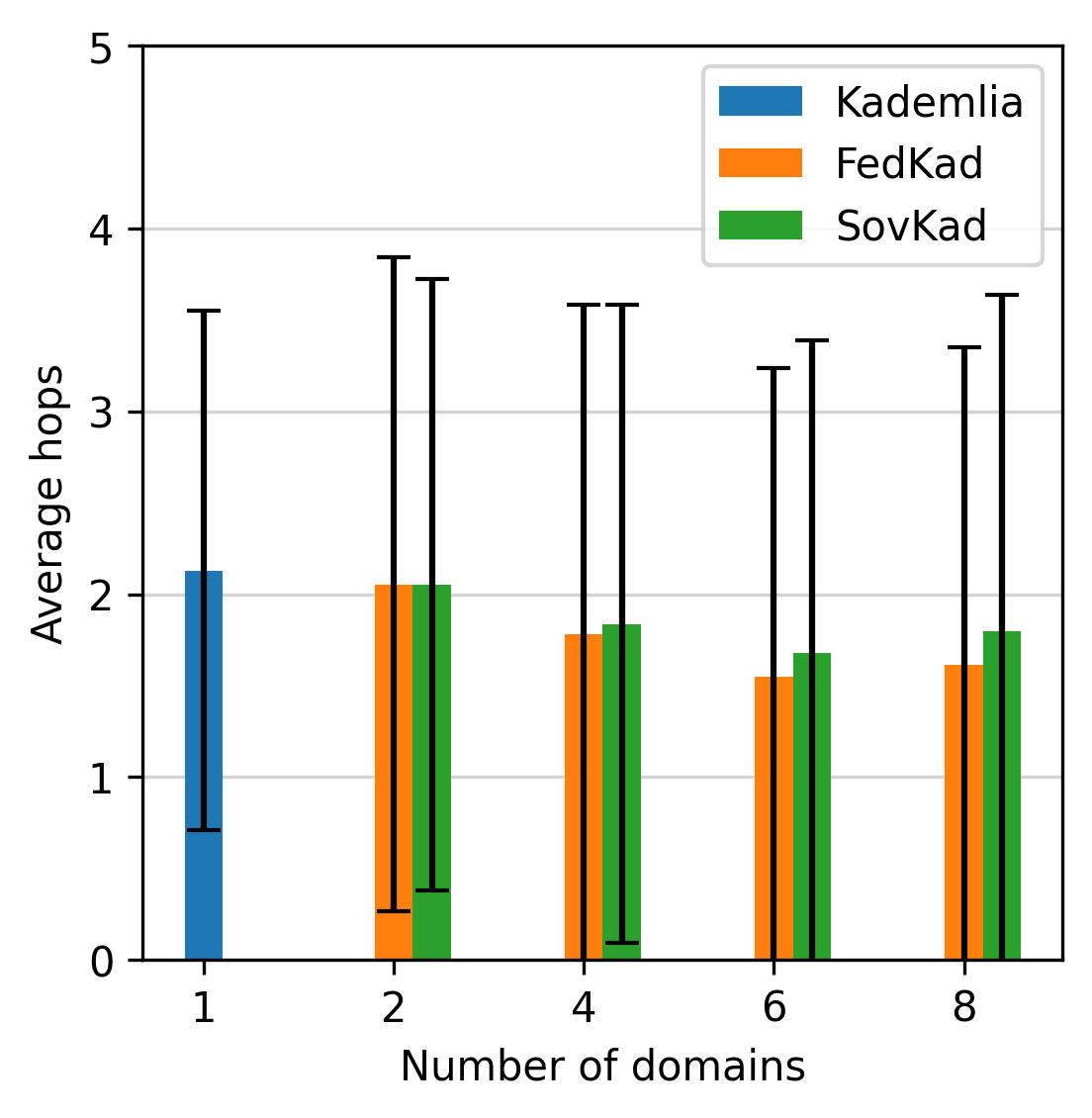}
         \caption{$|V|=8000$}
         \label{fig:byzantine_congregated_hops_8k}
     \end{subfigure}
     \hfill
     \begin{subfigure}[b]{0.3\textwidth}
         \centering
         \includegraphics[width=\textwidth]{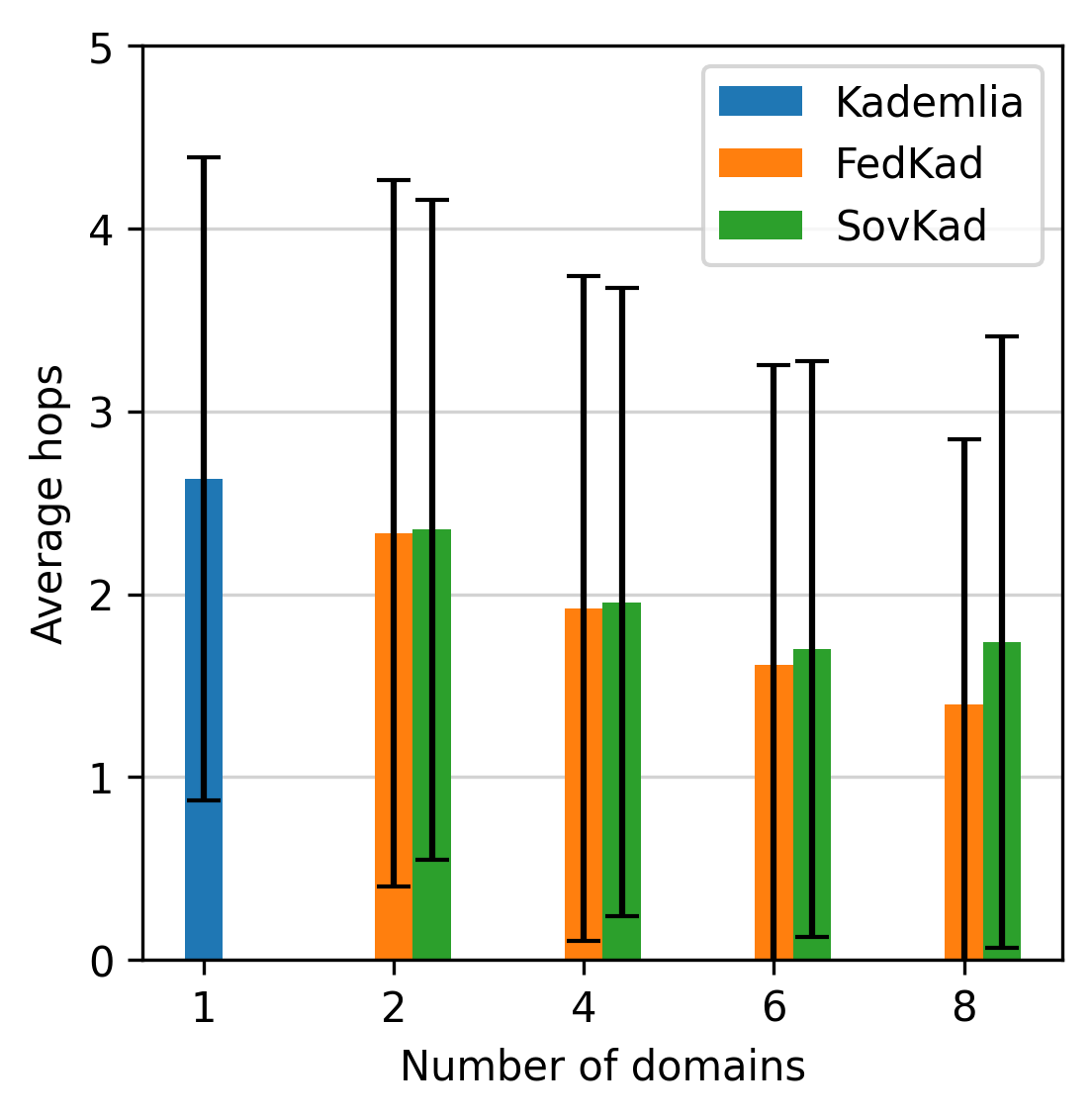}
         \caption{$|V|=16000$}
         \label{fig:byzantine_congregated_hops_16k}
     \end{subfigure}
        \caption{Average success routing path length of a successful lookup in a single overlay (Kademlia), the attacked domain in a federated multi-overlay (FedKad), and the attacked domain in a PODS (SovKad). The analysis is conducted for a scenario with  $f=0.3$, where all colluder nodes clustered in one of the domains. Bars represent standard deviation.}
        \label{fig:congregate-hops}
\end{figure}

\begin{figure}
     \centering
       \captionsetup[subfigure]{justification=centering}
     \begin{subfigure}[b]{0.3\textwidth}
         \centering
         \includegraphics[width=\textwidth]{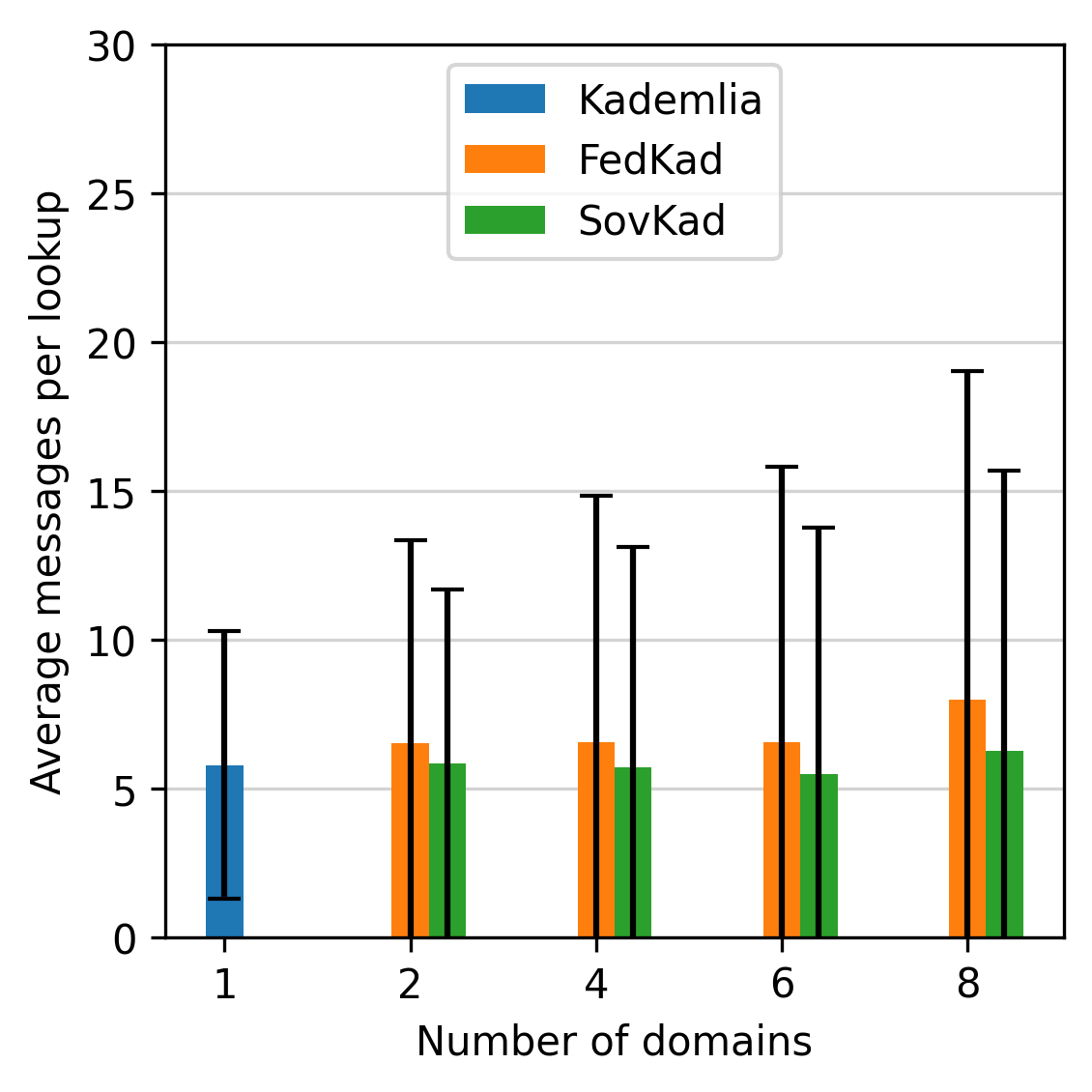}
         \caption{$|V|=1000$}
         \label{fig:byzantine_congregated_messages_1k}
     \end{subfigure}
     \hfill
     \begin{subfigure}[b]{0.3\textwidth}
         \centering
         \includegraphics[width=\textwidth]{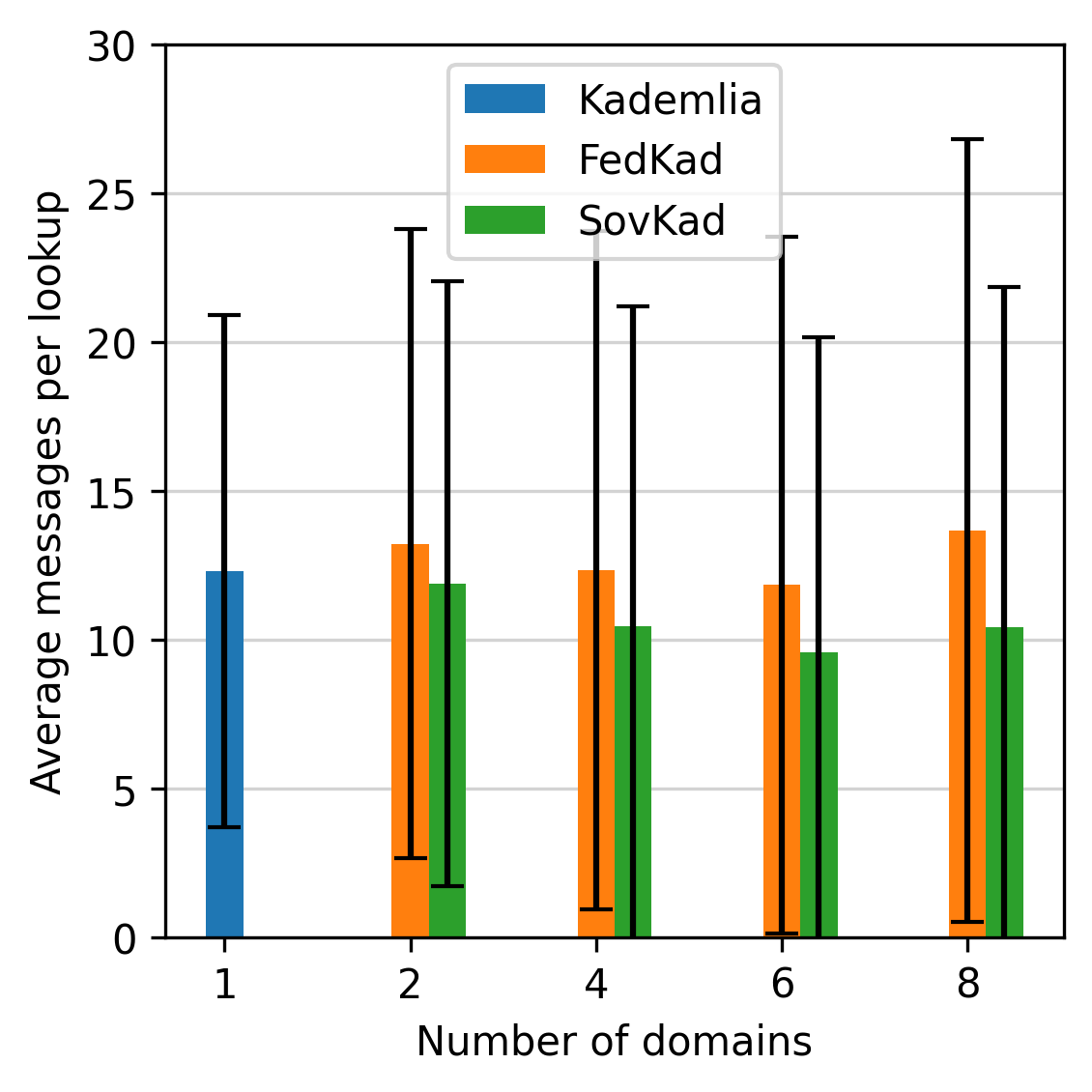}
         \caption{$|V|=8000$}
         \label{fig:byzantine_congregated_messages_8k}
     \end{subfigure}
     \hfill
     \begin{subfigure}[b]{0.3\textwidth}
         \centering
         \includegraphics[width=\textwidth]{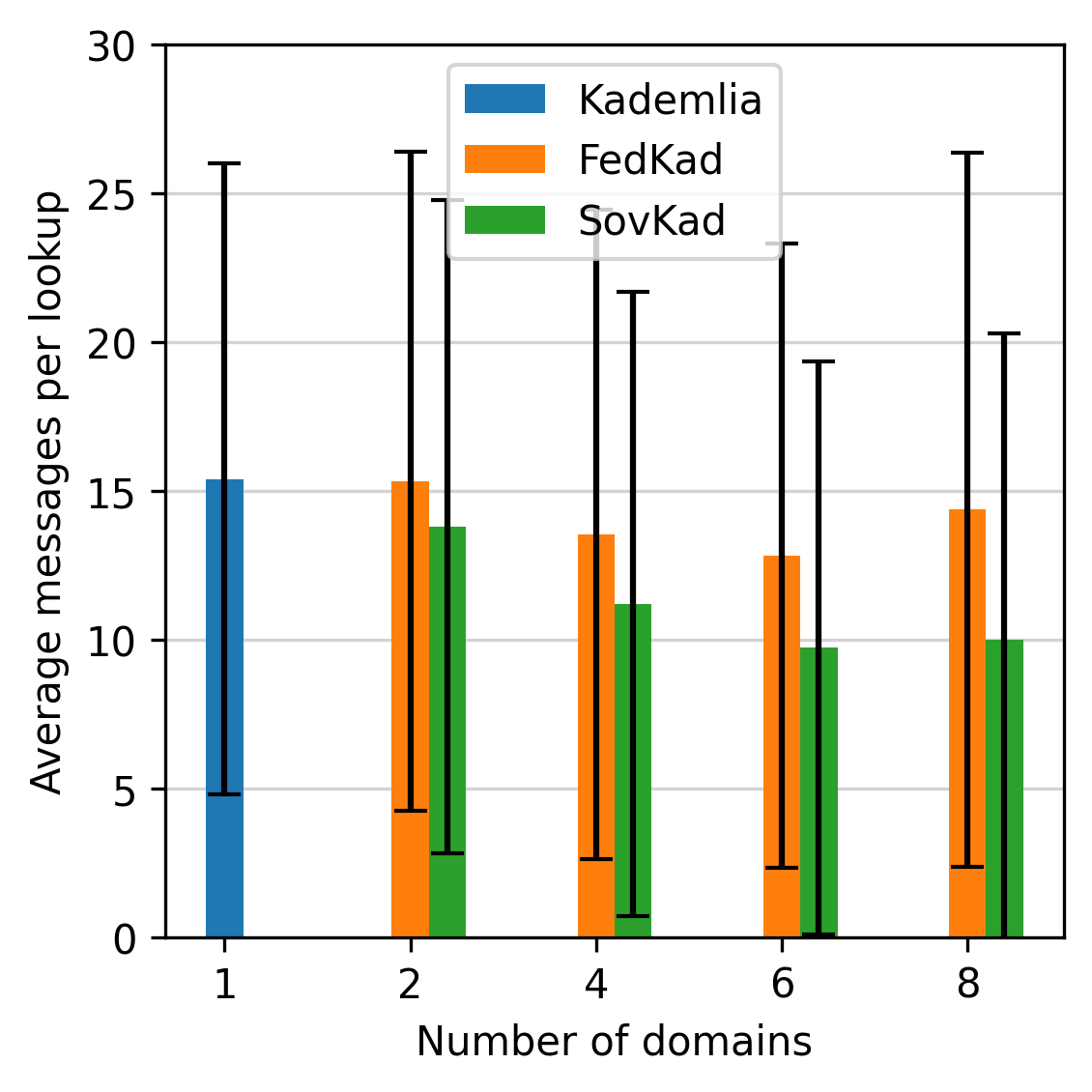}
         \caption{$|V|=16000$}
         \label{fig:byzantine_congregated_messages_16k}
     \end{subfigure}
        \caption{Average bandwidth cost per lookup, measured as the total number of messages sent,  in a single overlay (Kademlia), the attacked domain in a federated multi-overlay (FedKad), and the attacked domain in a PODS (SovKad). The analysis is conducted for a scenario with  $f=0.3$, where all colluder nodes are clustered in a single domain. Bars represent standard deviation.}
        \label{fig:congregate-bandwidth}
\end{figure}

\paragraph{Congregated Colluder Nodes}
Finally, we tested an attack in which the adversary tries to disrupt a specific domain, and moves all of its colluder nodes to that domain. 
Here again,  30\% of total nodes collude with the adversary ($f=0.3$).
Here the attacker is able to substantially damage FedKad lookup rates, relative to the Happy Path or the Uniform Distribution (see~\cref{fig:congregate-success}).
The attacked FedKad domain is flooded with colluder nodes, and has no ability to recover. 
The failing lookups require several retries, substantially increasing FedKad's bandwidth cost (see~\cref{fig:congregate-bandwidth}).

Using the active attack response mechanism, SovKad substantially outperforms FedKad, both in terms of success rate (\cref{fig:congregate-success}) and bandwidth cost (\cref{fig:congregate-bandwidth}). 
For both FedKad and SovKad, the Congregated Colluder attack slightly increases average path length for successful lookups (see~\cref{fig:congregate-hops}), but SovKad's path length has increased more than FedKad's. 
This is related to the success rate: requests that require longer paths are more likely to fail in FedKad, and thus not be counted toward the average.
SovKad's defense is not perfect: there is still a measurable difference between the Uniform Distribution and Congregated Colluder Node cases, but it is substantially less vulnerable to the Congregated attack than FedKad. 
These results demonstrate a key advantage of PODS: it enables attack responses that use the security of the network as a whole, rather than limiting security with the resources of each individual domain. 
\section{Conclusion}
\label{sec:conclusion}

We proposed a new scalable P2P overlay architecture for blockchain systems: P2P Overlay with Domain Sovereignty (PODS). To our knowledge, PODS is the first design that can accommodate node heterogeneity
without sacrificing performance or security. Our approach segments the overlay into independent sovereign domains, which have their own internal structure and set of protocols. These protocols can be tailored to the characteristics and needs of the nodes in the domain.

The PODS architecture combines the advantages of efficient information dissemination within a domain, similar to a smaller P2P overlay, with the security benefits of a larger P2P overlay. This is achieved by designing smaller domains within the architecture, enabling efficient communication within each domain, while also maintaining links between nodes from different domains. Additionally, the architecture incorporates an active response mechanism to address overlay attacks, further enhancing its security. A group-based reputation model is employed to detect and mitigate attacks, allowing nodes to detect and join the attacked domain for recovery purposes.

To demonstrate the effectiveness of the PODS architecture, we designed and implemented two novel node discovery protocols: FedKad and SovKad. Through simulations, we have compared the performance of SovKad with traditional node discovery methods such as Kademlia in a single overlay and FedKad in a federated multi-overlay scenario. Our results demonstrate that SovKad outperforms these existing approaches, providing empirical evidence of the potential of the PODS architecture in blockchain systems.



\bibliography{lipics-v2021-sample-article}

\begin{thebibliography}{10}

\bibitem{9524210}
Salem Alqahtani and Murat Demirbas.
\newblock Bottlenecks in blockchain consensus protocols.
\newblock In {\em 2021 IEEE International Conference on Omni-Layer Intelligent
  Systems (COINS)}, pages 1--8, 2021.
\newblock \href {https://doi.org/10.1109/COINS51742.2021.9524210}
  {\path{doi:10.1109/COINS51742.2021.9524210}}.

\bibitem{antwi2022survey}
Robert Antwi, James~Dzisi Gadze, Eric~Tutu Tchao, Axel Sikora, Henry
  Nunoo-Mensah, Andrew~Selasi Agbemenu, Kwame Opunie-Boachie Obour~Agyekum,
  Justice~Owusu Agyemang, Dominik Welte, and Eliel Keelson.
\newblock A survey on network optimization techniques for blockchain systems.
\newblock {\em Algorithms}, 15(6):193, 2022.

\bibitem{asgaonkar2022scaling}
Aditya Asgaonkar.
\newblock Scaling blockchains and the case for ethereum.
\newblock In {\em Handbook on Blockchain}, pages 197--213. Springer, 2022.

\bibitem{baumgart2007s}
Ingmar Baumgart and Sebastian Mies.
\newblock S/kademlia: A practicable approach towards secure key-based routing.
\newblock In {\em 2007 International conference on parallel and distributed
  systems}, pages 1--8. IEEE, 2007.

\bibitem{ipfs}
Juan Benet.
\newblock Ipfs - content addressed, versioned, p2p file system.
\newblock 07 2014.

\bibitem{bi2018accelerated}
Wei Bi, Huawei Yang, and Maolin Zheng.
\newblock An accelerated method for message propagation in blockchain networks.
\newblock {\em arXiv preprint arXiv:1809.00455}, 2018.

\bibitem{polkadotdesign}
Jeff Burdges, Alfonso Cevallos, Peter Czaban, Rob Habermeier, Syed Hosseini,
  Fabio Lama, Handan~Kilinc Alper, Ximin Luo, Fatemeh Shirazi, Alistair
  Stewart, et~al.
\newblock Overview of polkadot and its design considerations.
\newblock {\em arXiv preprint arXiv:2005.13456}, 2020.

\bibitem{narwhal}
George Danezis, Lefteris Kokoris-Kogias, Alberto Sonnino, and Alexander
  Spiegelman.
\newblock Narwhal and tusk: A dag-based mempool and efficient bft consensus.
\newblock In {\em Proceedings of the Seventeenth European Conference on
  Computer Systems}, EuroSys '22, page 34–50, New York, NY, USA, 2022.
  Association for Computing Machinery.
\newblock \href {https://doi.org/10.1145/3492321.3519594}
  {\path{doi:10.1145/3492321.3519594}}.

\bibitem{das2020efficient}
Sourav Das, Vinith Krishnan, and Ling Ren.
\newblock Efficient cross-shard transaction execution in sharded blockchains.
\newblock {\em arXiv preprint arXiv:2007.14521}, 2020.

\bibitem{surveynetworking}
Maya Dotan, Yvonne-Anne Pignolet, Stefan Schmid, Saar Tochner, and Aviv Zohar.
\newblock Survey on blockchain networking: Context, state-of-the-art,
  challenges.
\newblock {\em ACM Comput. Surv.}, 54(5), may 2021.
\newblock \href {https://doi.org/10.1145/3453161} {\path{doi:10.1145/3453161}}.

\bibitem{dotan2021survey}
Maya Dotan, Yvonne-Anne Pignolet, Stefan Schmid, Saar Tochner, and Aviv Zohar.
\newblock Survey on blockchain networking: Context, state-of-the-art,
  challenges.
\newblock {\em ACM Computing Surveys (CSUR)}, 54(5):1--34, 2021.

\bibitem{gervais2016security}
Arthur Gervais, Ghassan~O Karame, Karl W{\"u}st, Vasileios Glykantzis, Hubert
  Ritzdorf, and Srdjan Capkun.
\newblock On the security and performance of proof of work blockchains.
\newblock In {\em Proceedings of the 2016 ACM SIGSAC conference on computer and
  communications security}, pages 3--16, 2016.

\bibitem{HENDRIKX2015184}
Ferry Hendrikx, Kris Bubendorfer, and Ryan Chard.
\newblock Reputation systems: A survey and taxonomy.
\newblock {\em Journal of Parallel and Distributed Computing}, 75:184--197,
  2015.
\newblock URL:
  \url{https://www.sciencedirect.com/science/article/pii/S0743731514001464},
  \href {https://doi.org/https://doi.org/10.1016/j.jpdc.2014.08.004}
  {\path{doi:https://doi.org/10.1016/j.jpdc.2014.08.004}}.

\bibitem{bitcoinchurn}
Muhammad~Anas Imtiaz, David Starobinski, Ari Trachtenberg, and Nabeel Younis.
\newblock Churn in the bitcoin network: Characterization and impact.
\newblock In {\em 2019 IEEE International Conference on Blockchain and
  Cryptocurrency (ICBC)}, pages 431--439, 2019.
\newblock \href {https://doi.org/10.1109/BLOC.2019.8751297}
  {\path{doi:10.1109/BLOC.2019.8751297}}.

\bibitem{josang2007survey}
Audun J{\o}sang, Roslan Ismail, and Colin Boyd.
\newblock A survey of trust and reputation systems for online service
  provision.
\newblock {\em Decision support systems}, 43(2):618--644, 2007.

\bibitem{KOUTROULI201247}
Eleni Koutrouli and Aphrodite Tsalgatidou.
\newblock Taxonomy of attacks and defense mechanisms in p2p reputation
  systems—lessons for reputation system designers.
\newblock {\em Computer Science Review}, 6(2):47--70, 2012.
\newblock URL:
  \url{https://www.sciencedirect.com/science/article/pii/S1574013712000093},
  \href {https://doi.org/https://doi.org/10.1016/j.cosrev.2012.01.002}
  {\path{doi:https://doi.org/10.1016/j.cosrev.2012.01.002}}.

\bibitem{kwon2019cosmos}
Jae Kwon and Ethan Buchman.
\newblock Cosmos whitepaper.
\newblock {\em A Netw. Distrib. Ledgers}, page~27, 2019.

\bibitem{lavoie2017xor}
Erick Lavoie, Miguel Correia, and Laurie Hendren.
\newblock Xor-overlay topology management beyond kademlia.
\newblock In {\em 2017 IEEE 11th International Conference on Self-Adaptive and
  Self-Organizing Systems (SASO)}, pages 51--60. IEEE, 2017.

\bibitem{LI2020841}
Xiaoqi Li, Peng Jiang, Ting Chen, Xiapu Luo, and Qiaoyan Wen.
\newblock A survey on the security of blockchain systems.
\newblock {\em Future Generation Computer Systems}, 107:841--853, 2020.
\newblock URL:
  \url{https://www.sciencedirect.com/science/article/pii/S0167739X17318332},
  \href {https://doi.org/https://doi.org/10.1016/j.future.2017.08.020}
  {\path{doi:https://doi.org/10.1016/j.future.2017.08.020}}.

\bibitem{lin2009exploiting}
Shen Lin, Fran{\c{c}}ois Ta{\"\i}ani, and Gordon Blair.
\newblock Exploiting synergies between coexisting overlays.
\newblock In {\em IFIP International Conference on Distributed Applications and
  Interoperable Systems}, pages 1--15. Springer, 2009.

\bibitem{liu2022asymptotically}
Chen-Da Liu-Zhang, Christian Matt, and S{\o}ren~Eller Thomsen.
\newblock Asymptotically optimal message dissemination with applications to
  blockchains.
\newblock {\em Cryptology ePrint Archive}, 2022.

\bibitem{lopez2019please}
Pedro~Garcia Lopez, Alberto Montresor, and Anwitaman Datta.
\newblock Please, do not decentralize the internet with (permissionless)
  blockchains!
\newblock In {\em 2019 IEEE 39th International Conference on Distributed
  Computing Systems (ICDCS)}, pages 1901--1911. IEEE, 2019.

\bibitem{mast2022building}
Kai Mast.
\newblock Building protocols for scalable decentralized applications.
\newblock In {\em Handbook on Blockchain}, pages 215--255. Springer, 2022.

\bibitem{kademlia}
Petar Maymounkov and David Mazieres.
\newblock Kademlia: A peer-to-peer information system based on the xor metric.
\newblock In {\em International Workshop on Peer-to-Peer Systems}, pages
  53--65. Springer, 2002.

\bibitem{medrano2015performance}
Ad{\'a}n~G Medrano-Ch{\'a}vez, Elizabeth P{\'e}rez-Cort{\'e}s, and Miguel
  Lopez-Guerrero.
\newblock A performance comparison of chord and kademlia dhts in high churn
  scenarios.
\newblock {\em Peer-to-Peer Networking and Applications}, 8(5):807--821, 2015.

\bibitem{peersim}
Alberto Montresor and Mark Jelasity.
\newblock Peersim: A scalable p2p simulator.
\newblock In {\em 2009 IEEE Ninth International Conference on Peer-to-Peer
  Computing}, pages 99--100, 2009.
\newblock \href {https://doi.org/10.1109/P2P.2009.5284506}
  {\path{doi:10.1109/P2P.2009.5284506}}.

\bibitem{morris2003chord}
Robert Morris, M~Frans Kaashoek, David Karger, Hari Balakrishnan, Ion Stoica,
  David Liben-Nowell, and Frank Dabek.
\newblock Chord: A scalable peer-to-peer look-up protocol for internet
  applications.
\newblock {\em IEEE/ACM Transactions On Networking}, 11(1):17--32, 2003.

\bibitem{naik2020next}
Ashika~R Naik and Bettahally~N Keshavamurthy.
\newblock Next level peer-to-peer overlay networks under high churns: a survey.
\newblock {\em Peer-to-Peer Networking and Applications}, 13(3):905--931, 2020.

\bibitem{nakamoto2008bitcoin}
Satoshi Nakamoto.
\newblock Bitcoin: A peer-to-peer electronic cash system.
\newblock {\em Decentralized business review}, page 21260, 2008.

\bibitem{neudecker2015simulation}
Till Neudecker, Philipp Andelfinger, and Hannes Hartenstein.
\newblock A simulation model for analysis of attacks on the bitcoin
  peer-to-peer network.
\newblock In {\em 2015 IFIP/IEEE International Symposium on Integrated Network
  Management (IM)}, pages 1327--1332. IEEE, 2015.

\bibitem{8456488}
Till Neudecker and Hannes Hartenstein.
\newblock Network layer aspects of permissionless blockchains.
\newblock {\em IEEE Communications Surveys \& Tutorials}, 21(1):838--857, 2019.
\newblock \href {https://doi.org/10.1109/COMST.2018.2852480}
  {\path{doi:10.1109/COMST.2018.2852480}}.

\bibitem{free2shard}
Ranvir Rana, Sreeram Kannan, David Tse, and Pramod Viswanath.
\newblock Free2shard: Adversary-resistant distributed resource allocation for
  blockchains.
\newblock 6(1), feb 2022.
\newblock \href {https://doi.org/10.1145/3508031} {\path{doi:10.1145/3508031}}.

\bibitem{kadcast}
Elias Rohrer and Florian Tschorsch.
\newblock Kadcast: A structured approach to broadcast in blockchain networks.
\newblock In {\em Proceedings of the 1st ACM Conference on Advances in
  Financial Technologies}, AFT '19, page 199–213, New York, NY, USA, 2019.
  Association for Computing Machinery.
\newblock \href {https://doi.org/10.1145/3318041.3355469}
  {\path{doi:10.1145/3318041.3355469}}.

\bibitem{roos2015determining}
Stefanie Roos, Hani Salah, and Thorsten Strufe.
\newblock Determining the hop count in kademlia-type systems.
\newblock 2015.

\bibitem{shen2010handbook}
Xuemin Shen, Heather Yu, John Buford, and Mursalin Akon.
\newblock {\em Handbook of peer-to-peer networking}, volume~34.
\newblock Springer Science \& Business Media, 2010.

\bibitem{sit2002security}
Emil Sit and Robert Morris.
\newblock Security considerations for peer-to-peer distributed hash tables.
\newblock In {\em International Workshop on Peer-to-Peer Systems}, pages
  261--269. Springer, 2002.

\bibitem{srivatsa2006large}
Mudhakar Srivatsa, Bugra Gedik, and Ling Liu.
\newblock Large scaling unstructured peer-to-peer networks with
  heterogeneity-aware topology and routing.
\newblock {\em IEEE Transactions on Parallel and Distributed Systems},
  17(11):1277--1293, 2006.

\bibitem{tarkoma2010overlay}
Sasu Tarkoma.
\newblock {\em Overlay Networks: Toward Information Networking.}
\newblock CRC Press, 2010.

\bibitem{ethP2P}
Ethereum Team.
\newblock Networking layer, 2022.
\newblock URL:
  \url{https://ethereum.org/en/developers/docs/networking-layer/#devp2p}.

\bibitem{wang2019sok}
Gang Wang, Zhijie~Jerry Shi, Mark Nixon, and Song Han.
\newblock Sok: Sharding on blockchain.
\newblock In {\em Proceedings of the 1st ACM Conference on Advances in
  Financial Technologies}, pages 41--61, 2019.

\bibitem{yu2020survey}
Guangsheng Yu, Xu~Wang, Kan Yu, Wei Ni, J~Andrew Zhang, and Ren~Ping Liu.
\newblock Survey: Sharding in blockchains.
\newblock {\em IEEE Access}, 8:14155--14181, 2020.

\bibitem{rapidchain}
Mahdi Zamani, Mahnush Movahedi, and Mariana Raykova.
\newblock Rapidchain: Scaling blockchain via full sharding.
\newblock In {\em Proceedings of the 2018 ACM SIGSAC Conference on Computer and
  Communications Security}, CCS '18, page 931–948, New York, NY, USA, 2018.
  Association for Computing Machinery.
\newblock \href {https://doi.org/10.1145/3243734.3243853}
  {\path{doi:10.1145/3243734.3243853}}.

\bibitem{4577789}
Huanyu Zhao and Xiaolin Li.
\newblock H-trust: A robust and lightweight group reputation system for
  peer-to-peer desktop grid.
\newblock In {\em 2008 The 28th International Conference on Distributed
  Computing Systems Workshops}, pages 235--240, 2008.
\newblock \href {https://doi.org/10.1109/ICDCS.Workshops.2008.96}
  {\path{doi:10.1109/ICDCS.Workshops.2008.96}}.

\bibitem{zheng2017overview}
Zibin Zheng, Shaoan Xie, Hongning Dai, Xiangping Chen, and Huaimin Wang.
\newblock An overview of blockchain technology: Architecture, consensus, and
  future trends.
\newblock In {\em 2017 IEEE international congress on big data (BigData
  congress)}, pages 557--564. Ieee, 2017.

\end{thebibliography}

\appendix

\section{Pseudocodes}

\begin{algorithm}
\caption{Initiate FedKad Inter-Domain Lookup}
\DontPrintSemicolon
\SetKwFunction{FFindNode}{FindNode}
\SetKwProg{Fn}{Function}{:}{}
  \Fn{\FFindNode{\textit{source\_node}, \textit{target\_node}}}{
\textit{gateway\_node} $\gets $ selected uniformly at random from gateway nodes known to \textit{source\_node}\ (i.e., in xDHT)\;
Send a REQUEST message to \textit{gateway\_node} with target \textit{target\_node}\;
    Increase number of lookup attempts with 1

}
\label{algo:interFedKad}
\end{algorithm}

\begin{algorithm}
\caption{FedKad Gateway Node Handles REQUEST Message}
\DontPrintSemicolon
\SetKwFunction{FHandleFindNodeMessage}{HandleFindNodeMessage}
\SetKwProg{Fn}{Function}{:}{}
  \Fn{\FHandleFindNodeMessage{\textit{source\_node}, \textit{target\_node}}}{
    Add \textit{source\_node} to corresponding k-bucket in xDHT for domain of \textit{source\_node}\;
    \eIf{\textit{target\_node} in xDHT of target domain}{
      Send a RESPONSE message to \textit{source\_node} with target \textit{target\_node}\;
    }{
      $v_r\gets$closest node to target \textit{target\_node}\;
      Send a REQUEST message to $v_r$ with target \textit{target\_node}\;
    }
}
\label{algo:interGatewayFedKad}
\end{algorithm}

\begin{algorithm}
\caption{FedKad Domain Node Handles REQUEST Message from Gateway Node}
\DontPrintSemicolon
\SetKwFunction{FHandleFindNodeMessage}{HandleFindNodeMessage}
\SetKwProg{Fn}{Function}{:}{}
  \Fn{\FHandleFindNodeMessage{\textit{gateway\_node}, \textit{target\_node}}}{
    Add \textit{gateway\_node} to corresponding k-bucket in xDHT for gateway overlay\;
    \eIf{\textit{target\_node} in iDHT}{
      Send a RESPONSE message to \textit{gateway\_node} with target \textit{target\_node}\;
    }{
      $C\gets$ $\alpha$ closest nodes to target \textit{target\_node}\;
      Send a REQUEST message to $ \alpha$ \textit{candidate\_node} $\in C$ with target \textit{target\_node}\;
      Mark each candidate node \textit{candidate\_node} we sent a request to as visited\;
    }
}
\label{algo:interGatewayFedKadRequest}
\end{algorithm}

\begin{algorithm}
\caption{FedKad Domain Node Handles REQUEST Message from Domain Node}
\DontPrintSemicolon
\SetKwFunction{FHandleFindNodeMessage}{HandleFindNodeMessage}
\SetKwProg{Fn}{Function}{:}{}
  \Fn{\FHandleFindNodeMessage{\textit{domain\_node}, \textit{target\_node}}}{
    Add \textit{domain\_node} to corresponding k-bucket in iDHT\;
    \eIf{\textit{target\_node} in iDHT}{
      Send a RESPONSE message to \textit{domain\_node} with target \textit{target\_node}\;
    }{
      $C'\gets$ $\beta$ closest nodes to target \textit{target\_node}\;
      Send a RESPONSE message to \textit{domain\_node} with payload $C'$\;
    }
}
\label{algo:interDomainNodeFedKadRequest}
\end{algorithm}

\begin{algorithm}
\caption{FedKad Domain Node Handles RESPONSE Message from Domain Node}
\DontPrintSemicolon
\SetKwFunction{FHandleResponseMessage}{HandleResponseMessage}
\SetKwProg{Fn}{Function}{:}{}
  \Fn{\FHandleResponseMessage{\textit{domain\_node}, \textit{C'}}}{
    Add \textit{domain\_node} to corresponding k-bucket in iDHT\;
    \eIf{\textit{target\_node} in \textit{C'}}{
      Add \textit{target\_node} to \textit{C}\;
      Send a RESPONSE message to \textit{gateway\_node} with payload \textit{C}\;
    }{
      $C\gets$ $\beta$ nodes from \textit{C'}\;
      \eIf{\textit{C} has a node that is not marked as visited}{
        Send a REQUEST message to unvisited node \textit{domain\_node} with target \textit{target\_node}
        Mark \textit{domain\_node} as visited
      }{
        Return (search operation terminated)
      }
    }
}

\label{algo:interDomainNodeResponse}
\end{algorithm}

\begin{algorithm}
\caption{FedKad Domain Node Handles RESPONSE Message from Gateway Node}
\DontPrintSemicolon
\SetKwFunction{FHandleResponseMessage}{HandleResponseMessage}
\SetKwProg{Fn}{Function}{:}{}
  \Fn{\FHandleResponseMessage{\textit{gateway\_node}, \textit{C}}}{
    Add \textit{gateway\_node} to corresponding k-bucket in xDHT\;
    \eIf{\textit{target\_node} in \textit{C}}{
      Return (search operation terminated)
    }{
      \eIf{number of attempts < 3}{
        Initiate a new lookup (see \cref{algo:interFedKad})
      }{
        Return (search operation terminated)
      }
    }
}
\label{algo:interDomainNodeResponse}
\end{algorithm}

\begin{algorithm}
\caption{Initiate SovKad Inter-Domain Lookup}
\DontPrintSemicolon
\SetKwFunction{FFindNode}{FindNode}
\SetKwProg{Fn}{Function}{:}{}
  \Fn{\FFindNode{\textit{source\_node}, \textit{target\_node}}}{
    $R\gets$ $\alpha$ closest nodes to target \textit{target\_node} in \textit{source\_node}s xDHT for target domain\;
    Send a REQUEST message to $ \alpha$ \textit{relayer\_node}s $\in R$\ with target \textit{target\_node}\;
    Mark each relayer node \textit{relayer\_node} we sent a request to as visited\;
    Increase number of lookup attempts with 1

}
\label{algo:initateinterdomainlookupSovKad}
\end{algorithm}

\begin{algorithm}
\caption{SovKad Node Handles REQUEST Message from Node in Another Domain}
\DontPrintSemicolon
\SetKwFunction{FHandleFindNodeMessage}{HandleFindNodeMessage}
\SetKwProg{Fn}{Function}{:}{}
  \Fn{\FHandleFindNodeMessage{\textit{source\_node}, \textit{target\_node}}}{
    Add \textit{source\_node} to corresponding k-bucket in xDHT for domain of \textit{source\_node}\;
    \eIf{\textit{target\_node} in iDHT}{
      Send a RESPONSE message to \textit{source\_node} with target \textit{target\_node}\;
    }{
      $C\gets$closest candidate node to target \textit{target\_node} in iDHT\;
      Send a REQUEST message to \textit{candidate\_node} $\in C$ with target \textit{target\_node}\;
      Mark candidate node \textit{candidate\_node} as visited\;
    }
}
\label{algo:interHandleRequestAnotherDomain}
\end{algorithm}

\begin{algorithm}
\caption{SovKad Node Handles REQUEST Message from Node in Same Domain}
\DontPrintSemicolon
\DontPrintSemicolon
\SetKwFunction{FHandleFindNodeMessage}{HandleFindNodeMessage}
\SetKwProg{Fn}{Function}{:}{}
  \Fn{\FHandleFindNodeMessage{\textit{node}, \textit{target\_node}}}{
    Add \textit{node} to corresponding k-bucket in iDHT\;
    \eIf{\textit{target\_node} in iDHT}{
      Send a RESPONSE message to \textit{domain\_node} with target \textit{target\_node}\;
    }{
      $C'\gets$ $\beta$ closest nodes to target \textit{target\_node}\;
      Send a RESPONSE message to \textit{node} with payload $C'$\;
    }
}
\label{algo:interHandleRequestSameDomain}
\end{algorithm}

\begin{algorithm}
\caption{SovKad Node Handles RESPONSE Message from Node in Same Domain}
\DontPrintSemicolon
\SetKwFunction{FHandleResponseMessage}{HandleResponseMessage}
\SetKwProg{Fn}{Function}{:}{}
  \Fn{\FHandleResponseMessage{\textit{node}, \textit{C'}}}{
    Add \textit{node} to corresponding k-bucket in iDHT\;
    \eIf{\textit{target\_node} in \textit{C'}}{
      Add \textit{target\_node} to \textit{C}\;
      Send a RESPONSE message to \textit{source\_node} with payload \textit{C}\;
    }{
      $C\gets$ $\beta$ nodes from \textit{C'}\;
      \eIf{\textit{C} has a node that is not marked as visited}{
        Send a REQUEST message to unvisited node \textit{node} with target \textit{target\_node}\;
        Mark \textit{node} as visited
      }{
        Return (search operation terminated)
      }
    }
}

\label{algo:interDomainNodeResponse}
\end{algorithm}

\begin{algorithm}
\caption{SovKad Node Handles RESPONSE Message from Node in Different Domain}
\DontPrintSemicolon
\SetKwFunction{FHandleResponseMessage}{HandleResponseMessage}
\SetKwProg{Fn}{Function}{:}{}
  \Fn{\FHandleResponseMessage{\textit{node}, \textit{C}}}{
    Add \textit{node} to corresponding k-bucket in xDHT\;
    \eIf{\textit{target\_node} in \textit{C}}{
      Add \textit{target\_node} to \textit{C}\;
      Update domain quality score of target domain\;
      Return (search operation terminated)
    }{
      \eIf{number of attempts < 9}{
        Initiate a new lookup  \cref{algo:initateinterdomainlookupSovKad} (note that 9 because the three request messages in \cref{algo:initateinterdomainlookupSovKad} are independent)
      }{
        Update domain quality score of target domain\;
              \If{new domain quality score < quality threshold }{
              move to attacked domain 
              }
        Return (search operation terminated)
      }
    }
}

\label{algo:interDomainNodeResult}
\end{algorithm}

\end{document}